\documentclass[12pt,noshowpacs,nofootinbib,notitlepage,amsmath,superscriptaddress]{revtex4-1}
\usepackage{setspace}
\allowdisplaybreaks
\usepackage{graphicx,color}
\usepackage[colorlinks=true,citecolor=blue,linkcolor=blue,urlcolor=blue]{hyperref}
\usepackage[charter]{mathdesign}
\usepackage{multirow}
\usepackage{enumerate}

\DeclareSymbolFontAlphabet{\mathcal}{symbols}
\DeclareSymbolFont{symbols}{OMS}{xmdcmsy}{m}{n}
\DeclareSymbolFont{largesymbols}{OMX}{xmdcmex}{m}{n}
\SetSymbolFont{symbols}{bold}{OMS}{xmdcmsy}{b}{n}

 \newcommand{\be}{\begin{equation}}
\newcommand{\ee}{\end{equation}}
\newcommand{\beq}{\begin{equation}}
\newcommand{\eeq}{\end{equation}}
\newcommand{\bea}{\begin{eqnarray}}
\newcommand{\eea}{\end{eqnarray}}

\begin{document}
\title{\color{blue}\Large Quantum nucleation of up-down quark matter \\ and astrophysical implications}
\author{Jing Ren}
\email{renjing@ihep.ac.cn}
\affiliation{Institute of High Energy Physics, Chinese Academy of Sciences, Beijing 100049, P. R. China}
\author{Chen Zhang}
\email{czhang@physics.utoronto.ca}
\affiliation{Department of Physics, University of Toronto, Toronto, Ontario, Canada  M5S1A7}
\affiliation{Department of Physics, University of Waterloo, Waterloo, Ontario, Canada N2L 3G1}
\begin{abstract}
Quark matter with only $u$ and $d$ quarks ($ud$QM) might be the ground state of baryonic matter at large baryon number $A>A_{\rm min}$. With $A_{\rm min}\gtrsim 300$, this has no direct conflict with the stability of ordinary nuclei.  An intriguing test of this scenario is to look for quantum nucleation of $ud$QM inside neutron stars due to their large baryon densities. In this paper, we study the transition rate of cold neutron stars to $ud$ quark stars ($ud$QSs) and the astrophysical implications, considering the relevant theoretical uncertainties and observational constraints. It turns out that a large portion of parameter space predicts an instantaneous transition, and so the observed neutron stars are mostly $ud$QSs. We find this possibility still viable under the recent gravitational wave and pulsar observations, although there are debates on its compatibility with some observations that involve complicated structure of quark matter. The tension could be partially relieved in the two-families scenario, where the high-mass stars ($M\gtrsim2 M_{\odot}$) are all $ud$QSs and the low-mass ones  ($M\sim1.4\, M_{\odot}$) are mostly hadronic stars. In this case, the slow transition of the low-mass hadronic stars points to a very specific class of hadronic models with moderately stiff EOSs, and $ud$QM properties are also strongly constrained. 
\end{abstract}
\maketitle
\section{Introduction}

Quark matter, a state consisting purely of quark and gluon degrees of freedom without confining into individual nucleons, is expected to form at high density or high temperature. Bodmer~\cite{Bodmer:1971we}, Witten~\cite{Witten} and Terazawa~\cite{Terazawa:1979hq}, on the other hand, hypothesized that quark matter with comparable numbers of $u, \,d, \,s$ quarks, also called strange quark matter (SQM), might be the ground state of baryonic matter at the zero temperature and pressure. However, the original proposals are based on the bag model that fails to model the flavor-dependent feedback of the quark gas on the QCD vacuum. In our recent study~\cite{HRZ2017}, with this being adequately included in a phenomenological quark-meson model, we demonstrated that $u, d$ quark mater ($ud$QM) is in general more stable than SQM, and it can be more stable than the ordinary nuclear matter when the baryon number $A$ is sufficiently large above $A_{\rm min}\gtrsim 300$. This lower bound for $A_{\rm min}$ ensures the stability of ordinary nuclei and helps to avoid a catastrophic conversion of our empirical world\footnote{A stability analysis of the finite-size $ud$QM over giant nuclei in supernovae matter was carried by Ref.~\cite{Iida:2020fyt}.}. It also implies that the new form of stable matter has a relatively large positive charge, with $Z\gtrsim 100$ for $ud$QM staying just beyond the periodic table. These high-electric-charge objects can be searched for by their large ionization effects.\footnote{There have been searches for SQM in the cosmic ray or in samples of ordinary matter~\cite{Burdin:2014xma}. AMS in the space in particular has a great potential to identify high-charge particles with $Z\gtrsim 100$~\cite{Sandweiss:2004bu}. Recently there is a collier searches for such high-charge objects by using the LHC data~\cite{Aad:2019pfm}.}
The consequence of $e^{+} e^{-}$ pair production for $ud$QM with a large charge has also been investigated recently~\cite{Xia:2020byy}. 

One important question for the stable $ud$QM scenario is the implications for the neutron star physics. In the conventional picture, astrophysical neutron stars are assumed to be mostly hadronic stars  (HSs) as described by one family of equations of state (EOS), where hyperons are expected to appear in the high density region. 
However, the discovery of heavy pulsars with large masses above $2\,M_\odot$~\cite{Demorest:2010bx,Antoniadis:2013pzd,Cromartie:2019kug} ruled out a large number of soft hadronic matter EOSs as predicted by the presence of hyperons in the interiors. This conflict, also referred to as ``the hyperon puzzle", motivates an alternative explanation of these heavy pulsars as being pure quark stars. 

This possibility has been extensively studied in the context of the SQM hypothesis~\cite{Drago:2013fsa,Zacchi:2015lwa,Bombaci:2016xuj,Zhou:2017pha,Burgio:2018yix}, while it is a more natural option for the stable $ud$QM scenario. On one hand, with an intrinsic smaller effective bag constant for $ud$QM, $ud$ quark stars ($ud$QSs) can satisfy the $2\, M_\odot$ constraint more easily than strange stars~\cite{HRZ2017, Zhang:2019mqb, Zhao:2019xqy}.
On the other hand, hadronic stars consist mainly of $u, d$ quarks, and heavy stars are expected to convert to $ud$QSs much faster with an enhanced quantum nucleation rate of $ud$QM. There are still two possibilities, depending on the transition time for low-mass stars.
If the transition rate becomes significantly slow as the mass decreases to $1.4\, M_{\odot}$, these low-mass stars can remain hadronic in the present universe, leading to the two-families scenario that quark stars and hadronic stars coexist~\cite{Drago:2013fsa}. If the transition is fast at all relevant masses, it points to the less considered possibility that all compact stars are quark stars~\cite{Xu:2007wd}. 
Probes of $ud$QM and $ud$QSs via gravitational wave observations were explored in \cite{Zhang:2019mqb,Wang:2019jze}.

In this paper, we conduct a comprehensive and systematic study for the above two possibilities in the context of stable $ud$QM scenario, taking into account various uncertainties on the hadronic matter and quark matter properties, and the most recent observational constraints. We start with discussion of $ud$QM and $ud$QSs properties in Sec.~\ref{sec:QMQS}, as motivated by our recent study~\cite{HRZ2017}. To a good approximation, these properties are determined by an effective bag constant and a surface tension, the ranges of which are closely related to the stability condition of $ud$QM. In Sec.~\ref{sec:transition}, by adopting the standard calculation formalism for quantum nucleation, we identify the hadronic matter and $ud$QM features essential for the determination of the transition rate. The relevance of these two possibilities then becomes clear. In Sec.~\ref{sec:twoscenarios}, we discuss these two possibilities and their astrophysical observations in detail, where different information on $ud$QM and hadronic matter properties can be inferred. We conclude in Sec.~\ref{sec:conc}. In Appendix~\ref{tidal}, we present the detailed calculations for the tidal deformability constraints with updated results for the recent event GW190425 from LIGO/Virgo~\cite{Abbott:2020uma}.
In the rest of the paper, we use the natural unit with $c=\hbar=k_B=1$.

\section{Properties of $ud$QM and $ud$QS}
\label{sec:QMQS}

The novel possibility that $ud$QM is actually the ground state of baryonic matter was explored in an effective theory of sub-GeV mesons in our recent paper~\cite{HRZ2017}. Assuming a linear signal model, we fixed the free parameters in the meson potential by the masses and decay widths of mesons.
In the presence of finite quark densities, the meson fields are pushed away from the vacuum along the least steep direction. As a result, the constituent quark masses are reduced and quark matter becomes energetically favorable. Due to the badly broken flavor symmetry in QCD, the potential shape around the vacuum is much stiffer along the strange direction than the non-strange one. The $u, d$ quark mass then drops first as the Fermi momentum gradually increases from small values. Within the viable parameter space, an intermediate Fermi momentum is found to minimize the bulk energy per baryon $\varepsilon\equiv E/A$, where $u, d$ quark mass already becomes negligible and the strange fraction remains zero. Thus, in contrast to the naive expectation from the bag model, $ud$QM is more stable than SQM after taking into account the flavor symmetry breaking in the potential energy.

Around the Fermi momentum that $\varepsilon$ is minimized, the energy per baryon for $ud$QM in the bulk limit (large baryon number $A\gg 1$) can be well approximated by contributions from  a relativistic quark gas and from a spatially constant potential energy,
\begin{eqnarray}\label{eq:epsilon1}
\varepsilon\equiv \frac{\rho}{n}\approx 
\frac{3}{4}\, N_C\, p_F\, \chi+\frac{3\pi^2 B_{\rm eff}}{p_F^3}\,,
\end{eqnarray}
where  $\rho$ is the energy density and $n$ is the baryon number density. $N_C=3$ is the color factor for quarks, $\chi=\sum_i f_i^{4/3}$ is the flavor factor with the fraction $f_i=n_i/(N_C\, n)$,  and $p_F=(3\pi^2 n)^{1/3}$ is the Fermi momentum. The effective bag constant $B_{\rm eff}$ denotes the potential difference along the valley oriented close to the non-strange direction. It is rather insensitive to $p_F$ and is closely related to the lightest meson mass. The minimum energy per baryon is then
\be
\varepsilon_{\rm min}\approx 3\sqrt{2\pi}\,\chi_0^{3/4}B_{\rm eff}^{1/4}\,,
\label{EperA_analy}
\ee
at the Fermi momentum $p_{F,0}\approx \sqrt{2\pi}\,\chi_0^{-1/4}B_{\rm eff}^{1/4}$, with $\chi_0=(2/3)^{4/3}+(1/3)^{4/3}$ for a charge neutral $u, d$ gas. This shows the direct connection between the energy per baryon and the effective bag constant.
For a large part of parameter space, we find $ud$QM in the bulk limit more stable than the most stable nuclei ${}^{56}$Fe, i.e. $\varepsilon_{\rm min}\lesssim 930\,$MeV,  and so it is the ground state of baryonic matter with zero pressure. 

Small $ud$QM becomes less stable due to the finite size effects and the Coulomb energy contribution. For the baryon number $A$ not too small, the former can be well approximated by a surface-tension term for the quark-vacuum interface. From the numerical fit, the surface tension is found to be quite insensitive to the variation of relevant parameters, with a robust value $\sigma_{s0}\approx 20\,\rm MeV\,fm^{-2}$. To not ruin the stability of ordinary nuclei, it is safe to have the minimum baryon number $A_\textrm{min}$ of $ud$QM larger than 300, corresponding to $\varepsilon_{\rm min}\gtrsim 900\,$MeV. Therefore, with Eq.~(\ref{EperA_analy}), the scenario of  stable $ud$QM predicts the range of the effective bag constant to be
\begin{eqnarray}\label{eq:Beffrange}
50\,\rm MeV\,fm^{-3} \lesssim B_{\rm eff}\lesssim 57\, \rm MeV\, fm^{-3}. 
\label{Bbound1}
\end{eqnarray}

The stable $ud$QM scenario could be realized in a more general setup. Going beyond the simple model in \cite{HRZ2017}, the upper bound on $B_{\rm eff}$ remains intact as it is directly related to the stability condition in the bulk limit. The lower bound derived from the condition $A_\textrm{min}\gtrsim 300$, on the other hand, could be relaxed if the effective surface tension is larger, as predicted in some other models~\cite{Wen:2010zz}. Assuming the same Coulomb contribution in the analytical approximation for the energy, a more general lower bound is  $\sigma_{s0}/({\rm MeV\,fm^{-2}})+2\,B_{\rm eff}/({\rm MeV\,fm^{-3}})\gtrsim 120$, e.g. $\sigma_{s0}\gtrsim 30\,\rm MeV\,fm^{-2}$ is required for $B_{\rm eff}\approx 45\, \rm MeV\,fm^{-3}$. As for SQM, a large perturbative QCD effect or a color superconducting phase could reduce $B_{\rm eff}$ to a smaller range for the same stability condition of $\varepsilon_{\rm min}$~\cite{Burgio:2018yix, Zhou:2017pha, Weissenborn:2011qu}.

The physics of compact stars relies on the properties of hadronic matter and $ud$QM at certain temperature and pressure~\cite{Bombaci:2016xuj,Most:2019onn,Dexheimer:2019pay}. In this paper, we restrict to cold stars with zero temperature as a good approximation for mature neutron stars being formed after some time. 
The approximation Eq.~(\ref{eq:epsilon1}) ceases to apply at large pressure when a nonzero strange fraction becomes favored. It turns out that the pressure within reach for stable $ud$QSs remains small and the strange fraction can be safely ignored, as we will show at the end of this section. Thus,  in the rest of the paper, we stick with the effective bag constant range (\ref{eq:Beffrange}) inferred from the $ud$QM properties. 
Quantum nucleation of $ud$QM in hadronic matter phase relies on quark matter and hadronic matter properties at the same pressure. The flavor composition of $ud$QM is then determined by that of the hadronic matter in chemical equilibrium due to the conservation of baryon and lepton numbers, and differs from $ud$QM in equilibrium. Depending on the models, there could be a considerable fraction of electrons and muons in a neutron star interior, corresponding to an increasing number of protons.

To derive $ud$QM properties as functions of the pressure, we start from $ud$QM with relativistic electrons. The muon contribution is corrected by the non-negligible mass, and it will be discussed later. The energy density $\rho$ as a function of $n$ and $f_i$ can be found from Eq.~(\ref{eq:epsilon1}) with $N_C=1$ for leptons,
\begin{eqnarray}\label{eq:rhon}
\rho\approx \frac{9}{4}(3\pi^2)^{1/3}\,\chi\, n^{4/3}+B_{\rm eff}\,,\quad
\chi=\left(\frac{2}{3}-\frac{1}{3}f_p\right)^{4/3}+\left(\frac{1}{3}+\frac{1}{3}f_p\right)^{4/3}+\frac{1}{3}f_e^{4/3}\,.
\end{eqnarray}
$f_e, f_p$ denote the electron and proton fractions of the hadronic matter, and $f_e=f_p$ when muons are absent. 
With Eq.~(\ref{eq:rhon}), the pressure can be found through the thermodynamic relation
\be
P=n^2\left.\frac{\partial ({\rho/n})}{\partial n}\right|_{f_i}
\approx \frac{1}{3}\left(\rho-\rho_0\right)\,,
\label{Prho}
\ee
where $\rho_0=\varepsilon_{\rm min}n_0\approx 4B_{\rm eff}$ is the surface density with zero pressure. Combining these two equations, we obtain the baryon number density 
\be
n(P)\approx\frac{2}{3}\sqrt{\frac{2}{{\pi}}}\,\chi^{-3/4} \left(P+B_{\rm eff}\right)^{3/4}\,.
\label{nP_analy}
\ee
The chemical potential $\mu$ can be found by substituting Eq.~(\ref{Prho}) and Eq.~(\ref{nP_analy}) into another thermodynamics relation
\be
\mu(P)=\frac{\rho+P}{n}\approx3\sqrt{2\pi}\, \chi^{3/4}(P+B_{\rm eff})^{1/4}\,,
\label{muP_analy}
\ee
with $\mu(0)=\varepsilon_{\rm mim}$ at the surface. This expression is equivalent to $\mu=\sum_{i=u,d,e} N_C \,f_i \,\mu_i$, where $\mu_i=p_{Fi}$ for relativistic particles. For increasing electron fraction $f_e$, the flavor factor $\chi$ is larger, and $ud$QM formed via transition has a smaller $n(P)$ and a larger $\mu(P)$ in comparison to that in $\beta$-equilibrium. For $f_e\sim \mathcal{O}(10\%)$, the chemical potential can change by the order of 10\,MeV. 

Including muons, instead of a mere redefinition of $\chi$, there are non-negligible mass corrections to the energy density in Eq.~(\ref{eq:rhon}) with the muon mass comparable to $p_F$. 
For thermodynamic quantities relevant to quantum tunneling, the major change is for $\mu(P)$ with the additional contribution $f_\mu \mu_\mu$ in Eq.~(\ref{muP_analy}). Due to the chemical equilibrium, $\mu_\mu=\mu_e$, $f_\mu<f_e$, and the muon contribution is bounded from above by that from electrons.  

As a useful approximation, if the $P$ dependence of the flavor factor $\chi$ is mild, the thermodynamic relation in  Eq.~(\ref{Prho}) can be rewritten as $P\approx n^2d(\rho/n)/dn$. Together with Eq.~(\ref{muP_analy}), this leads to a simple relation $d\mu/dP\approx 1/n$ for either hadronic matter or quark matter. In the integral form, it becomes
\be
\label{eq:muPnP}
\mu(P)\approx \mu(0)+\int_0^P dP'\frac{1}{n(P')}\,.
\ee
As we will show in Sec.~\ref{sec:transition}, this relation is crucial in understanding the general feature of hadronic matter to $ud$QM transition.


Next, we discuss the properties of $ud$QSs. The crucial quantity is the EOS of $ud$QM in Eq.~(\ref{Prho}). It takes the same form as SQM in the bag model if ignoring the effect of strange quark mass, with the coefficient for $\rho$ expected for a relativistic gas and a non-vanishing surface density $\rho_0=4B_{\rm eff}$. Referring to Eq.~(\ref{EperA_analy}), $ud$QM with the same minimum energy par baryon $\varepsilon_{\rm min}$ as SQM has a smaller surface density $\rho_0$ due to a larger value of the flavor factor $\chi$. For quark stars with an enormous $A$, gravitational interaction becomes important, and the density profile can be found by solving  the Tolman-Oppenheimer-Volkoff (TOV) equation~\cite{Oppenheimer:1939ne,Tolman:1939jz}
\begin{equation}
{dP(r)\over dr}=-{G\left[m(r)+4\pi r^3P(r)\right]\left[\rho(r)+P(r)\right]\over
r(r-2Gm(r))}\,,\qquad{dm(r)\over dr}=4\pi\rho(r)r^2\,,
\label{eq:tov}
\end{equation}
with the $ud$QM EOS in (\ref{Prho}). As in the case for strange stars, this linear form of $ud$QM EOS enables one to
rewrite the TOV equation in terms of the following dimensionless variables~\cite{Zdunik:2000xx,Haensel:2007yy},
\be
\bar\rho=\frac{\rho}{\rho_0},\quad
\bar{P}=\frac{P}{\rho_0},\quad
\bar{r}=r\sqrt{G\rho_0},\quad
\bar{m}=m\sqrt{G^{3}\rho_0}\,,
\label{rescale}
\ee 
and the $\rho_0$ or $B_{\rm eff}$ dependence in the TOV equation is fully absorbed into the rescaled solution. 
 
 \begin{figure}[h]
\centering
\includegraphics[width=8cm]{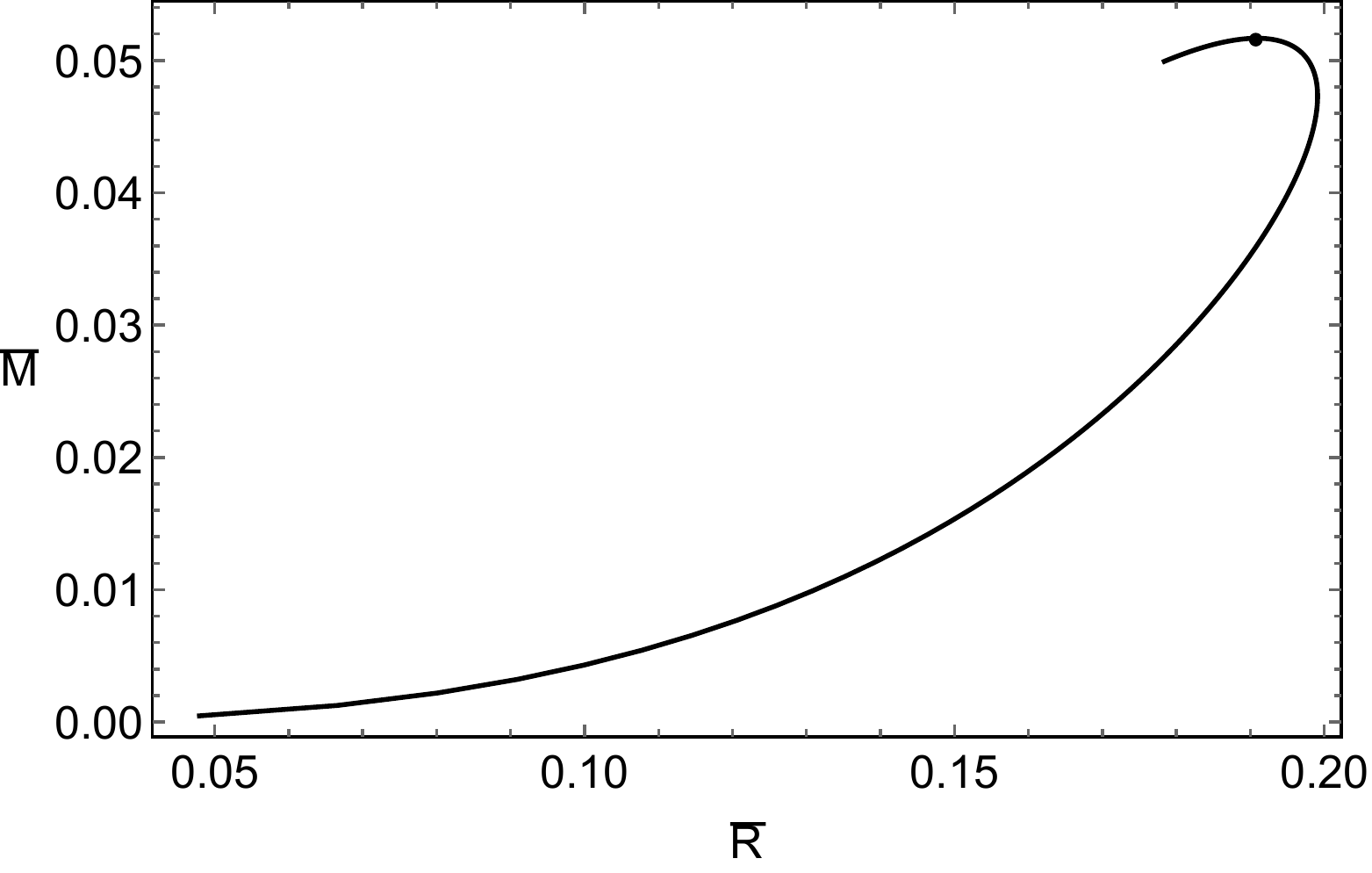}\quad  
\includegraphics[width=8cm]{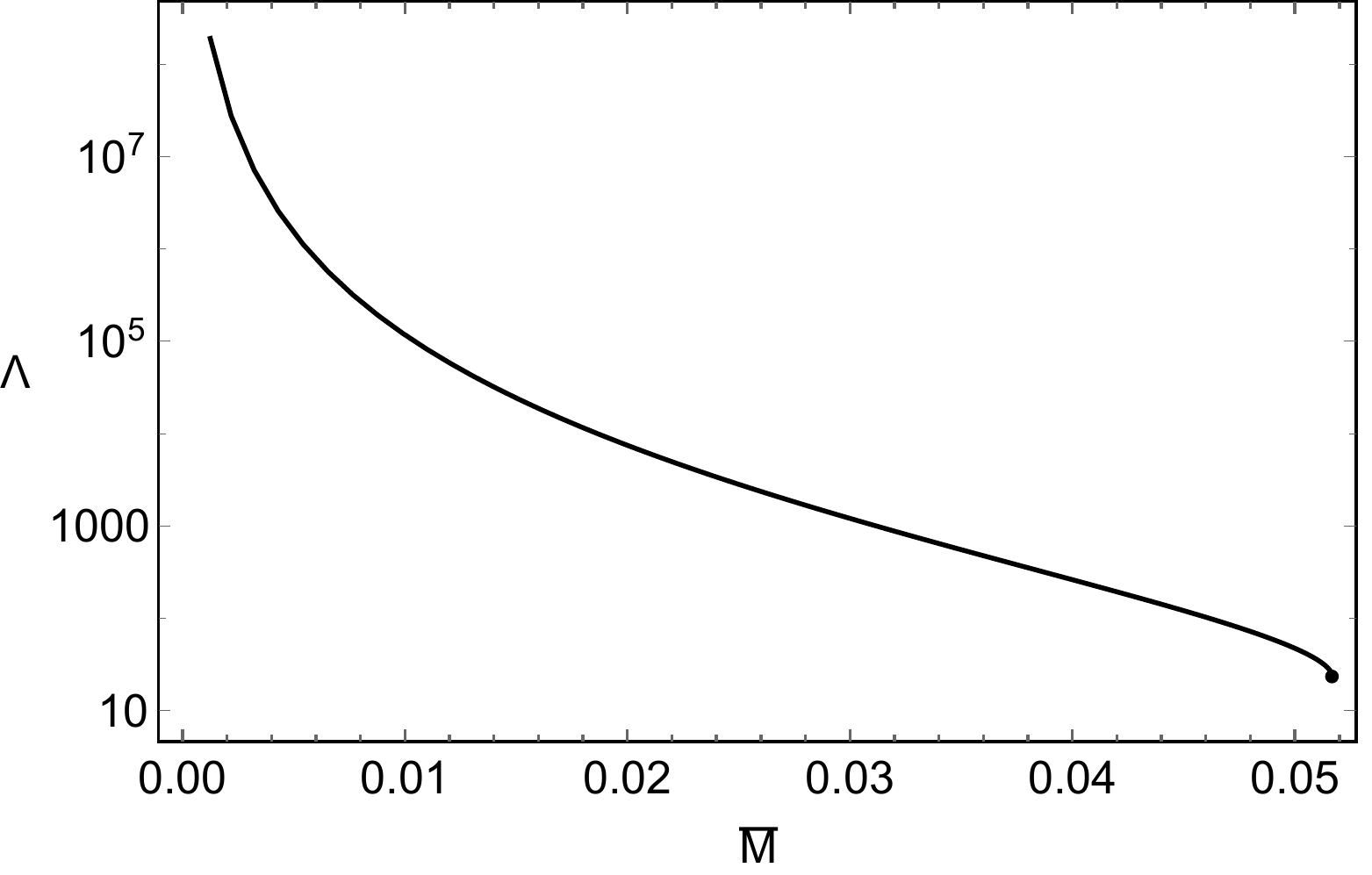}  
 \caption{ (a) Rescaled mass $\bar{M}$ vs rescaled radius $\bar{R}$  and (b) tidal deformability $\Lambda$ vs $\bar{M}$ for $ud$QSs. The black dot denotes the maximum mass configuration with $(\bar{M}, \bar{R})\approx (0.052, 0.19)$.} 
\label{fig:$ud$QS}
\end{figure}

Figure~\ref{fig:$ud$QS}\,(a) displays the mass-radius relation for the rescaled solution. Since quark matter is self-bound, the radius vanishes when the mass approaches zero. The stable branch of $ud$QSs extend all the way up to the maximum mass configuration with $\bar{M}_{\rm max}\approx0.052$, corresponding to $M_{\rm max}\approx15.2M_\odot (B_{\rm eff}/{\rm MeV\,fm^{-3}})^{-1/2}$.  The radio measurements of heavy pulsar masses around $2M_\odot$ provide a lower bound for $M_{\rm max}$ and then an upper bound for $B_{\rm eff}$. 
The most stringent upper bound comes from the recent observation of J0740+6620 with $M\approx 2.14^{+0.10}_{-0.09}\,M_{\odot}$~\cite{Cromartie:2019kug},\footnote{More massive pulsars have been suggested based on the optical spectroscopic and photometric observations~\cite{Linares:2018ppq}. But, as we will show later in Fig.~\ref{fig:BeffconsS1}, this imposes less stronger constraints due to large uncertainties of this method in comparison to the Shapiro delay measurements through radio timing.} indicating $B_{\rm eff}\lesssim50.3^{+4.5}_{-4.4}\,\rm MeV\, fm^{-3}$ at 68\%  confidence level~\cite{Zhang:2019mqb}. We can see that the $B_{\rm eff}$  range (\ref{eq:Beffrange}) in the stable $ud$QM scenario remains consistent with this upper bound. Due to uncertainties related to the hadronic matter and $ud$QM properties, the lower mass neutron stars with $M\sim 1.4M_\odot$ could be either hadronic stars or quark stars. For the latter case, i.e. all compact star being quark stars scenario, the radii of $ud$QSs with $M\sim 1.4M_\odot$ can be compared with observations. This then provides additional constraints on $B_{\rm eff}$, as we will discuss in Sec.~\ref{sec:scenario1}. 

Fig.~\ref{fig:$ud$QS}\,(b) shows the tidal deformability $\Lambda$ of $ud$QSs in terms of the rescaled mass. As a useful quantity to characterize the tidal properties of $ud$QS, the tidal deformability is determined by the Love number $k_2$ and the compactness $C=G M/R=\bar{M}/\bar{R}$ with $\Lambda=2k_2/(3C^5)$. As discussed in Appendix~\ref{tidal}, the dimensionless rescaling Eq.~(\ref{rescale}) can be extended to equations for $k_2$ so that the Love number for $ud$QSs is determined only by $C$ and is independent of $B_{\rm eff}$. We find $k_2$ ranging from 0.7 to 0.06 as the compactness increases. Given the $\bar{M}-\bar{R}$ relation, we can present $\Lambda$ as a function of $\bar{M}$. Since the compactness increases with the mass, the tidal deformability becomes small for heavy stars with strong gravitational interactions, and it reaches the minimum value $\Lambda_{\rm min}\approx 23$ at the maximum mass.  
The possibility that gravitational wave observations of coalescing neutron stars involve $ud$QSs has been discussed in~\cite{Zhang:2019mqb,Wang:2019jze} for GW170817 from LIGO/Virgo ~\cite{TheLIGOScientific:2017qsa}. In this paper, we extend the discussion to a newer event GW190425~\cite{Abbott:2020uma}, and constrain $B_{\rm eff}$ together with other observations considering the transition rate estimation of neutron stars.

As a final remark, we justify the earlier assumption of ignoring the strange fraction.  
The central pressure for the maximum mass $ud$QSs has the rescaling relation: $P_{\rm max}\approx1.3 \rho_0=5.1\, B_{\rm eff}$. From Eq. (\ref{nP_analy}), the corresponding Fermi momentum $p^{\rm max}_{\rm F}=(3 \pi^2 n_{\rm max})^{1/3}\approx 4.1 \, B_{\rm eff}^{1/4}$, which can reach up to  $590\,$MeV for $B_{\rm eff}$ in Eq.~(\ref{Bbound1}). In comparison to the Fermi momentum that nonzero strange fraction is favored, $p^{\rm max}_{\rm F}$ is larger only for about ten percent of models from our parameter scan~\cite{HRZ2017}.\footnote{The strange fraction will turn on above the special Fermi momentum $p_F^{(s)}$, when it is energetically favorable to produce non-relativistic or relativistic strange quarks. This gives $p_F^{(s)}\approx \textrm{min}\,[1.1m_s, 3.2\Delta V_s^{1/4}]$, where $m_s\gtrsim 550\,$MeV is the constituent mass for non-relativistic strange quark and $\Delta V_s\gtrsim 7.4\times 10^{8}\,{\rm MeV}^4$ is the potential energy change due to a shift of the fields.}  
Thus, for most of the parameter space, it is a reasonable assumption to ignore the strangeness at the pressure accessible from a stable quark star.


\section{Quantum nucleation of $ud$QM in cold neutron star matter}
\label{sec:transition}

\begin{figure}[h]
\centering
\includegraphics[width=10cm]{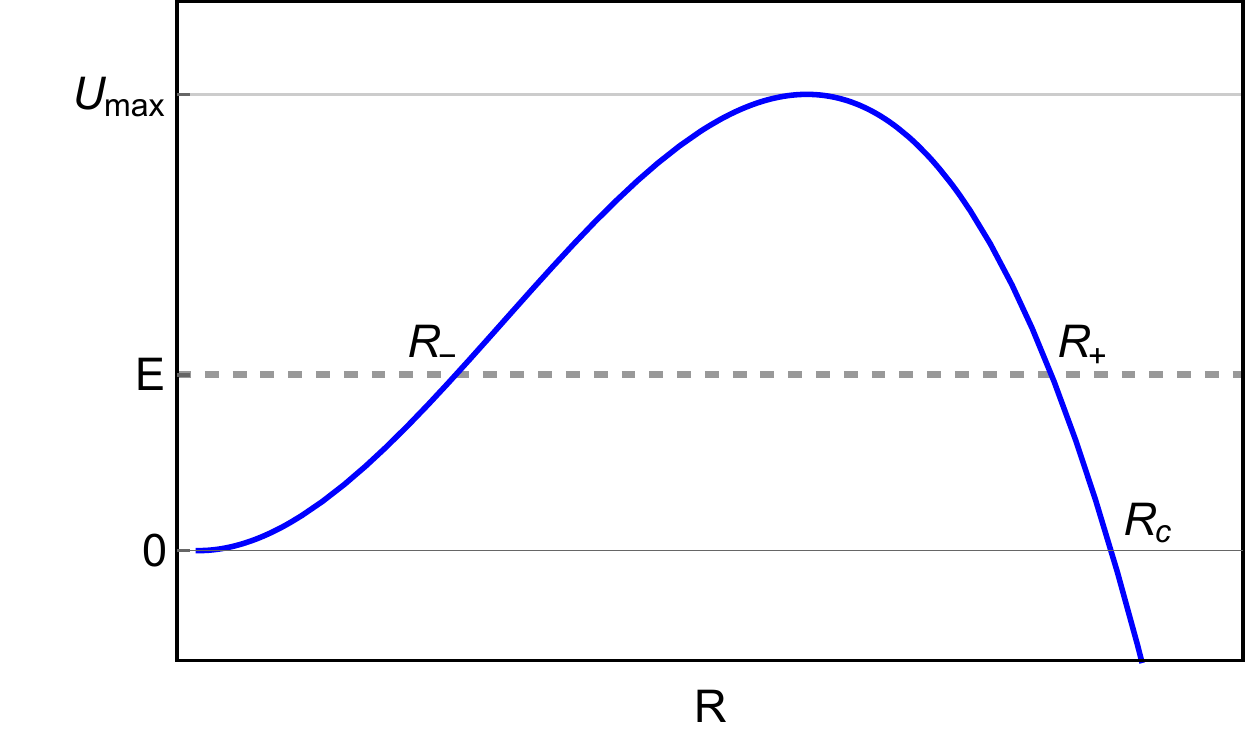} 
 \caption{The potential energy $U({\cal R})$ for a stable phase droplet with a radius ${\cal R}$. $U_{\rm max}$ denotes the potential peak value. ${\cal R}_c$ is the nontrivial zero of the potential and denotes the typical size of a droplet. ${\cal R}_\pm$ denote the classical turning points in the tunneling rate calculation of a state of energy $E$.}
\label{Ushape}
\end{figure}

We start by reviewing the calculation framework for the quark matter nucleation rate. 
Inside a cold neutron star, a droplet of more stable $ud$QM nucleates in the metastable hadronic phase through quantum tunneling. In the semiclassical approximation, the virtual droplet can be described by a sphere with radius ${\cal R}(t)$. The potential energy for such a fluctuation can be represented as~\cite{lk72,iida98}
\begin{equation}
  U({\cal R}) = \frac{4}{3}\pi n_{Q}(\mu_{Q} - \mu_H){\cal R}^3 + 4\pi \sigma_s {\cal R}^2
  \equiv -C_P{\cal R}^3 + 4\pi \sigma_s {\cal R}^2
\label{eq:potential} 
\end{equation}
where $C_P\equiv 4\pi n_{Q}(\mu_{H} - \mu_Q)/3$. $\mu_H$, $\mu_Q$ are the chemical potentials of hadron and quark matter, and $n_Q$ is the baryon number density of the later. $\sigma_s$ is the surface tension for the quark-hadron interface, and it differs from $\sigma_{s0}$ for the quark-vacuum interface in general. As $\sigma_s$ may suffer more from the theoretical uncertainties, i.e. $\sigma_s\sim 10$--$150 \rm \,MeV\,fm^{-2}$ found for different models in the literature,\footnote{Conventional MIT bag model~\cite{Berger:1986ps}, NJL model~\cite{Ke:2013wga,Garcia:2013eaa}, linear sigma model~\cite{Palhares:2010be,Pinto:2012aq,Fraga:2018cvr} predict small surface tension $\sigma_s\lesssim\,\rm30\, MeV\,fm^{-2}$. However, large values are also obtained, e.g. $\sigma_s\approx145\textrm{--}165 \rm\, MeV/fm^2$ for NJL model in the multiple reflection expansion framework~\cite{Lugones:2013ema} and $\sigma_s\approx 50\textrm{--}150 \rm \, MeV/fm^2$ for models including charge screening effects~\cite{Voskresensky:2002hu}.} we treat $\sigma_s$ as a free parameter in this paper.
For this potential, the first term denotes the negative volume contribution that favors the quark matter with $\mu_Q<\mu_H$, and the second term denotes  the positive surface contribution that prevents nucleation at smaller radii. A potential barrier forms due to competition of the two contributions, as shown in Fig.~\ref{Ushape}. It is useful to characterize the potential by its peak value and a special radius that denotes the typical size of a droplet, 
\begin{eqnarray}
U_{\rm max}=\frac{256\pi^3\sigma_s^3}{27 C_P^2},\quad
{\cal R}_c=\frac{4\pi\sigma_s}{ C_P}\,.
\end{eqnarray}

The kinetic energy of a droplet results from a flow in the medium around the droplet when there is a density discontinuity between the two phases. 
For a general case, the Lagrangian for the fluctuation can be written as~\cite{iida98}
\begin{eqnarray}
\mathcal{L}=M({\cal R})-M({\cal R})\sqrt{1-\dot{\cal R}^2}-U({\cal R})\,,
\end{eqnarray}
where $\dot{{\cal R}}$ is the growth rate and $M(R)$ is the effective mass for the droplet, 
\begin{eqnarray}\label{eq:mass}
M({\cal R})=4\pi \rho_H\left(1-\frac{n_Q}{n_H}\right)^2{\cal R}^3 \equiv C_M {\cal R}^3\,,
\end{eqnarray} 
where $C_M\equiv 4\pi \rho_H\left(1-n_Q/n_H\right)^2$ and $\rho_H$ is the energy density for the hadronic phase.  The kinetic term incorporates the relativistic effects. When $\dot{{\cal R}}\ll1$, it takes the non-relativistic form $\frac{1}{2}M({\cal R})\dot{\cal R}^2$ as in the Lifshitz-Kagan theory~\cite{LK}. 

For the quantum tunneling problem, a state of energy $E$ satisfies the Schrodinger equation, 
\begin{eqnarray}
\left[-\frac{d^2}{dR^2}+(U({\cal R})-E)(2M({\cal R})+E-U({\cal R}))\right]\psi({\cal R})=0\,.
\end{eqnarray}
With the standard semiclassical (WKB) approximation, the tunnelling probability for one droplet is
\begin{eqnarray}
p_0=\exp\left[-A(E_0)\right]\,.
\end{eqnarray}
In the non-relativistic limit, $A(E)$ is roughly the action under the potential barrier. Taking into account relativistic effects, it takes the form
\begin{equation}
A(E) = 2
\int_{{\cal R}_-}^{{\cal R}_+} d{\cal R}\sqrt{[2{\cal M}({\cal R}) +E-U({\cal R})][U({\cal R})-E]}\,.
\label{eq:AE}
\end{equation}
${\cal R}_\pm$ denote the classical turning points as given by $U({\cal R}_\pm)=E$. 
The ground state energy $E_0$ is determined from Bohr's quantization condition
\begin{eqnarray}
I(E_0)=2\pi \left(m_0+\frac{3}{4}\right),
\end{eqnarray} 
where $I(E)$ is the action for the zero-point oscillation,
\begin{eqnarray}
I(E)=2\int_0^{{\cal R}_-} d{\cal R}\sqrt{[2M({\cal R})+E-U({\cal R})][E-U({\cal R})]}\,.
\end{eqnarray}
$m_0=\left[I(E_\textrm{min})/(2\pi)+1/4\right]$, with [...] the Gauss' notation and $E_\textrm{min}=\max\,[U({\cal R})-2M({\cal R})]\propto \sigma_s^3/(C_P+2C_M)^2$ the minimum allowed energy for $E_0$ when the relativistic effects are large, i.e. $U({\cal R})>2M({\cal R})$. 
The transition time for one droplet is then,
\begin{eqnarray}\label{eq:tunrate}
\tau= \left(\nu_0\exp\left[-A(E_0)\right]\right)^{-1}\,,
\end{eqnarray}
where $\nu_0^{-1}=\left.dI/dE\right|_{E=E_0}$.

Inside a hadronic star, the transition time Eq.~(\ref{eq:tunrate})  for a $ud$QM droplet at the radius $r$ is determined by the properties of $ud$QM and hadronic matter evaluated at the pressure $P(r)$, through the coefficients $C_P, \sigma_s$ in the potential energy Eq.~(\ref{eq:potential}) and $C_M$ in the effective mass Eq.~(\ref{eq:mass}). After its quantum formation, the first droplet quickly expands by eating up nucleons and the whole star will be converted almost instantaneously. The transition time for a hadronic star $\tau_s$ can then be approximated by formation time of the first droplet, 
\begin{eqnarray}
\tau_s\approx \frac{\tau_\textrm{min}}{N_s}\,, 
\end{eqnarray}
where $\tau_\textrm{min}$ denotes the minimum transition time a droplet could have  inside the star and $N_s\gg 1$ denotes the number of such droplets. Given that the neutron star radius is around 10\,km and the typical size of a droplet is in the order of fm, we have roughly $N_s\sim (\textrm{km}/\textrm{fm})^3\sim 10^{54}$. A more careful estimate for transition at the core gives $N_s\sim 10^{48}$~\cite{iida98}. To account for the related uncertainties, we assume $N_s\sim 10^{45}$--$10^{55}$ in the rest of the paper. 
\begin{figure}[h]
\centering
\includegraphics[width=10cm]{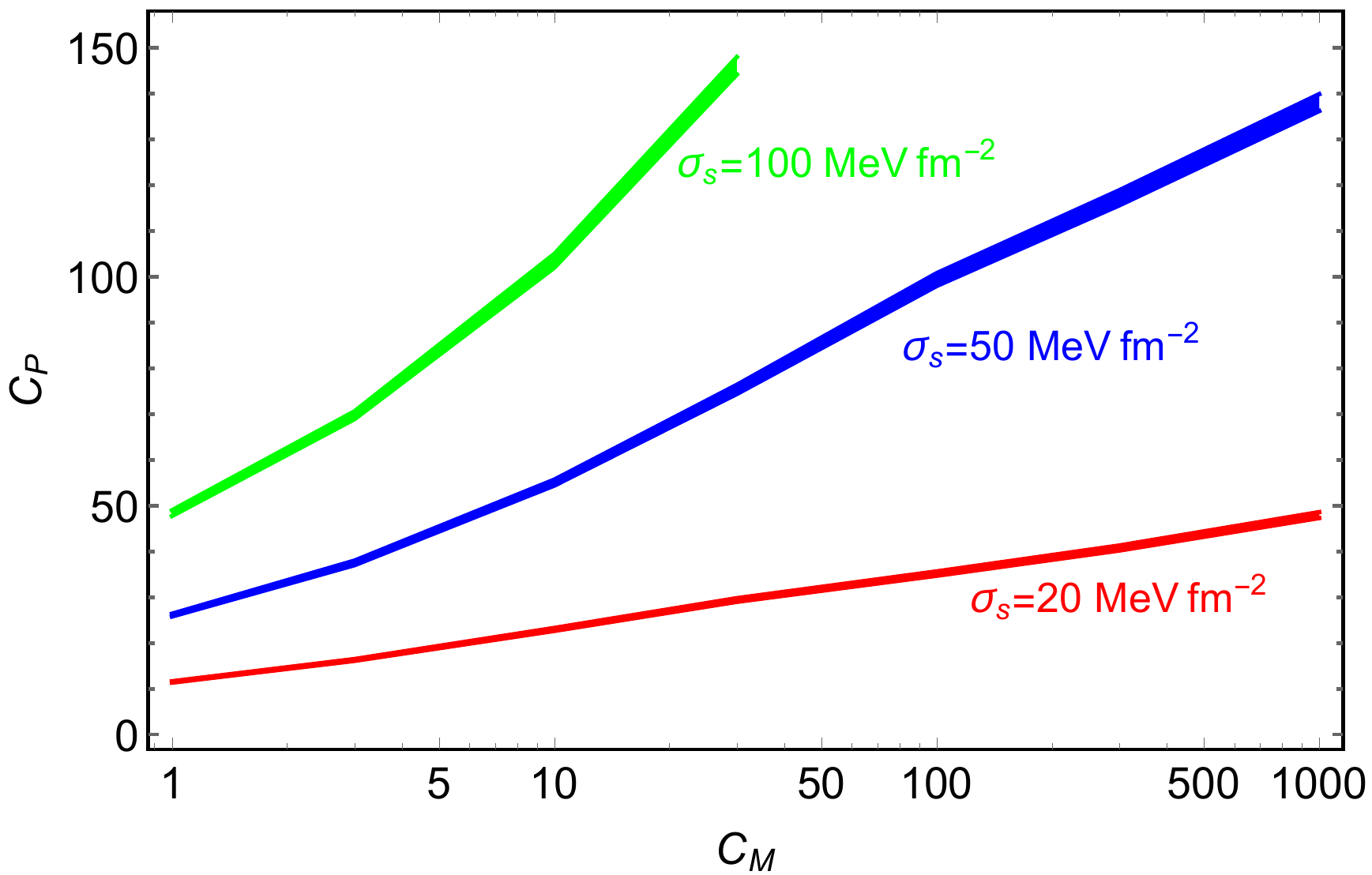}  
 \caption{Contours of the transition time for a hadronic star $\tau_s\approx t_0\approx 4\times 10^{17}\,$s (the age of the universe )  on the plane of $C_P=\frac{4}{3}\pi n_{Q}(\mu_{H} - \mu_Q)$ and $C_M=4\pi \rho_H\left(1-n_Q/n_H\right)^2$ for $N_s\sim 10^{45}$--$10^{55}$ and the surface tension $\sigma_s=20, 50, 100 \rm\, MeV\,fm^{-2}$.}
\label{fig:Cscan}
\end{figure}

Figure~\ref{fig:Cscan} shows the contours of $\tau_s$ being the age of the universe on the $C_P-C_M$ plane for a given surface tension for the quark-hadron interface. 
Considering the large uncertainties for $\sigma_s$, we show the contour for some different values within the plausible range. A larger $\sigma_s$ increases the height of the potential barrier, leading to a larger $A(E)$ in Eq.~(\ref{eq:AE}) and a smaller tunneling probability. A smaller effective mass $C_M$, on the other hand, lifts $E_{\rm min}$ and thus the lowest energy $E_0$, which then increases the tunneling probability. When $C_M$ approaches zero, corresponding to vanishing discontinuity of the number densities, $E_{\rm min}$ rises up to $U_0$, inducing an instantaneous transition. $C_P$ influences both the potential barrier and $E_0$. A decreasing $C_P$ lifts the potential barrier height more than its lifting of $E_{\rm min}$, and this makes the transition slower.

\begin{figure}[h]
\centering
\includegraphics[width=7.5cm]{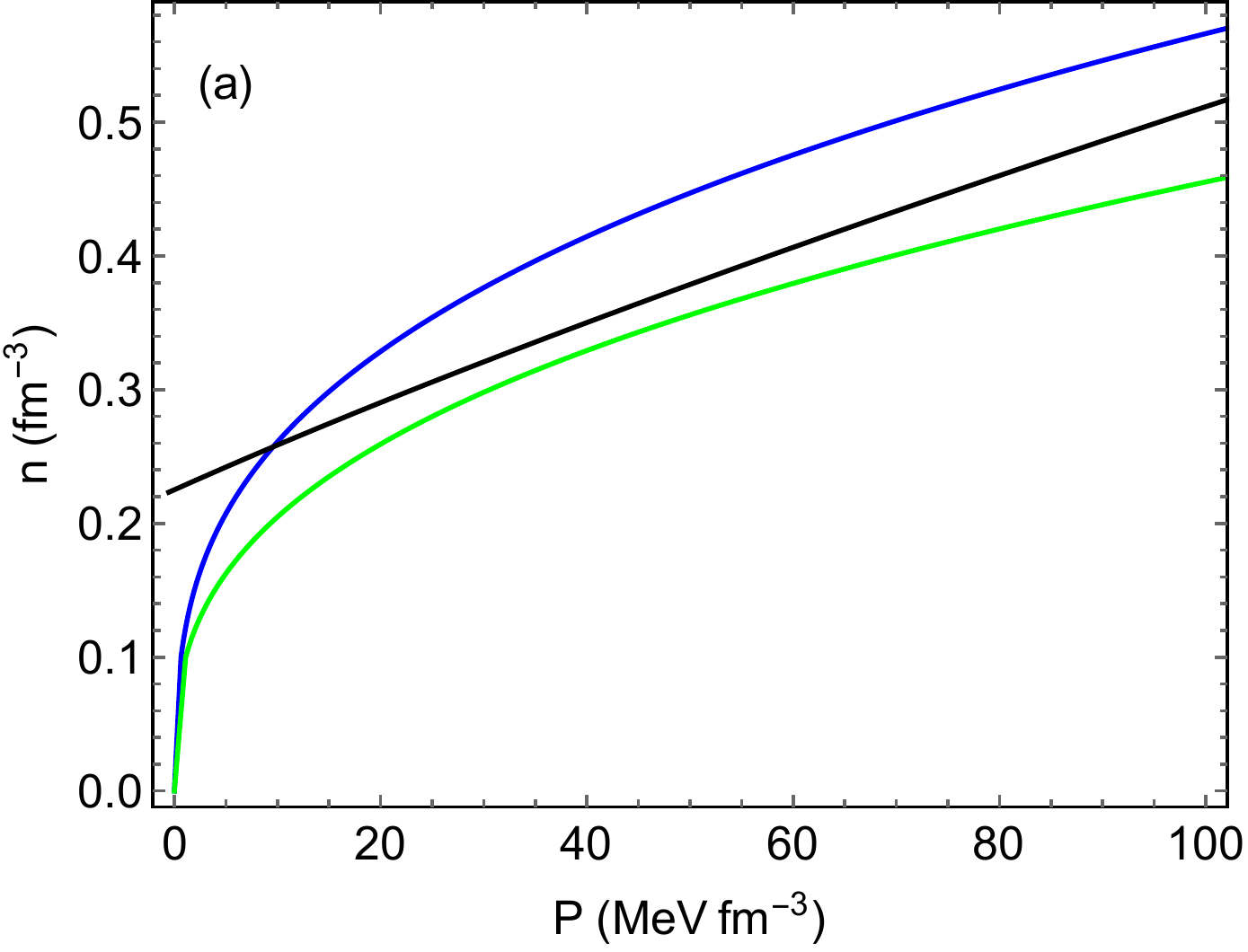}\quad  
\includegraphics[width=7.7cm]{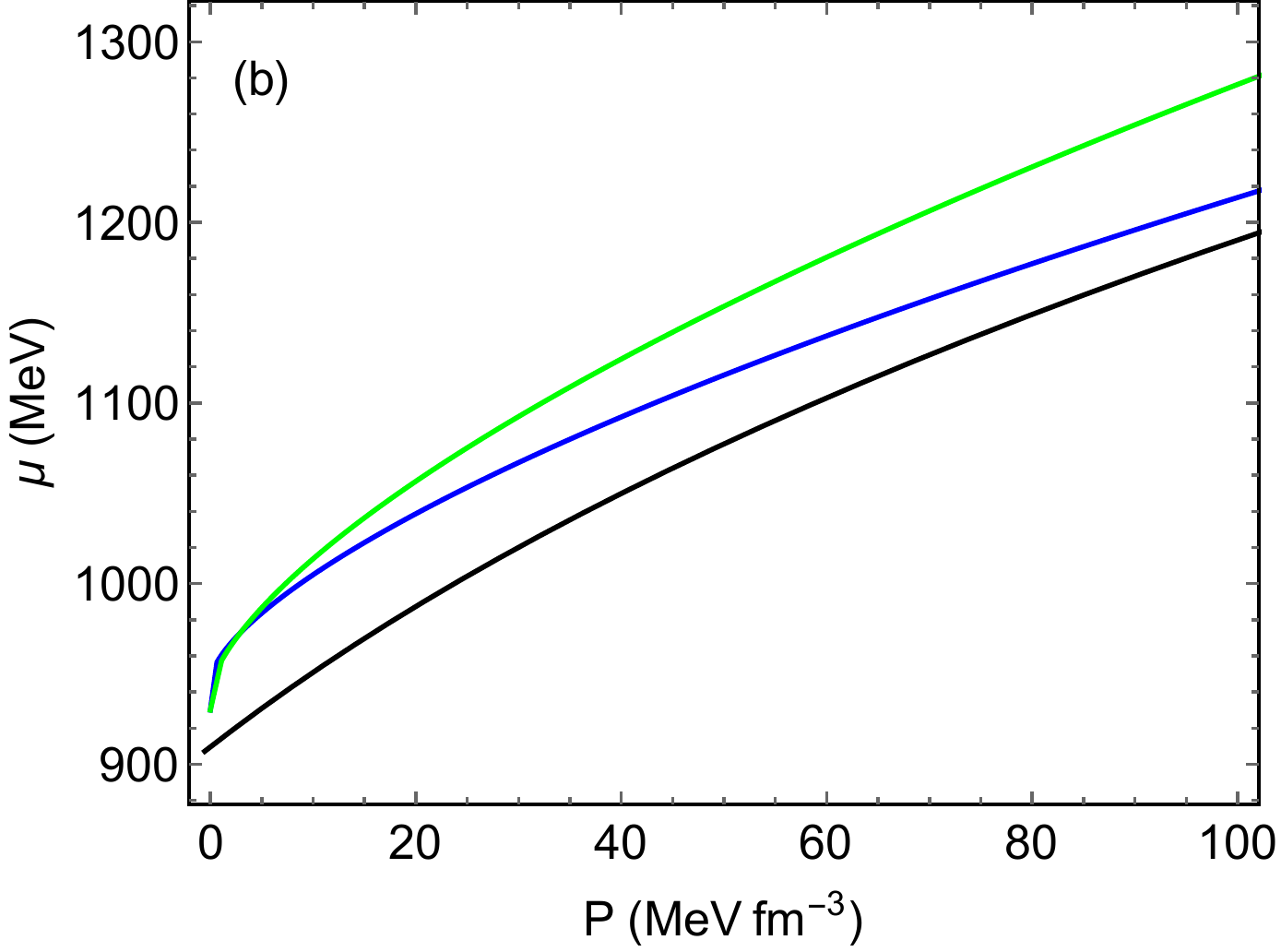}  
 \caption{ (a) Number density $n(P)$ and (b) chemical potential $\mu(P)$ for two examples of widely-used hadronic matter EOSs~\cite{Compose}, SLy (blue), GM1 (green), and $ud$QM with $B_{\rm eff}=52\,\rm MeV\,fm^{-3}$ (black).} 
\label{munPcompare}
\end{figure}

It turns out that the two important quantities $C_P\propto \mu_H-\mu_Q$ and $C_M\propto (n_Q-n_H)^2$ are closely related. Fig.~\ref{munPcompare} displays the chemical potential $\mu$ and the number density $n$ as functions of the pressure $P$ for some hadronic matter and $ud$QM models. 
The surface properties at zero pressure are more or less fixed. The chemical potential difference $\Delta \varepsilon_\textrm{min}\equiv \mu_H(0)-\mu_Q(0)\approx 930\,\textrm{MeV}-\varepsilon_\textrm{min}>0$ is directly related to the binding energy of $ud$QM, with $\mu_H(0)\approx 930$\,MeV for $^{56}{\rm Fe}$. For the density, $n_H(0)$ approximately vanishes for hadronic matter  in general, while $n_Q(0)$ is nonzero for self-bound $ud$QM as determined also by $\varepsilon_\textrm{min}$. The properties in the interior depend on the stiffness of the hadronic matter EOS. 
A soft hadronic EOS has the density increasing more rapidly with the pressure, e.g. SLy, and its $n(P)$ curve may intersect with the $ud$QM curve at a small pressure.  For such a case, $C_M$ approaches zero around the intersection radius and the transition is instantaneously fast regardless of the values for other quantities. The intersection of $n(p)$ curves can be avoided for a stiff hadronic EOS such as GM1, but the chemical potential difference and then $C_P$ also become larger for this case. The final result for the transition time depends on the competition between $C_P$ and $C_M$. This competition is expected due to the relation between the density difference and the chemical potential difference as from the thermodynamic relation in Eq.~(\ref{eq:muPnP}),
\be
\mu_H-\mu_Q\approx \Delta \varepsilon_{\rm min}+\int_0^P dP\left(\frac{1}{n_H}-\frac{1}{n_Q}\right)\,.
\label{mup_int}
\ee
Therefore, for the case that $n_H$ is bounded from above by $n_Q$, both $C_M$ and $C_P$ become larger for a stiffer hadronic matter EOS. At certain point, a too large $C_P$ dominates the transition time and the increasing stiffness would not help to slow down the transition.

\section{Conversion of neutron stars and astrophysical observations}
\label{sec:twoscenarios}

\begin{figure}[h]
\centering
\includegraphics[width=8.0cm]{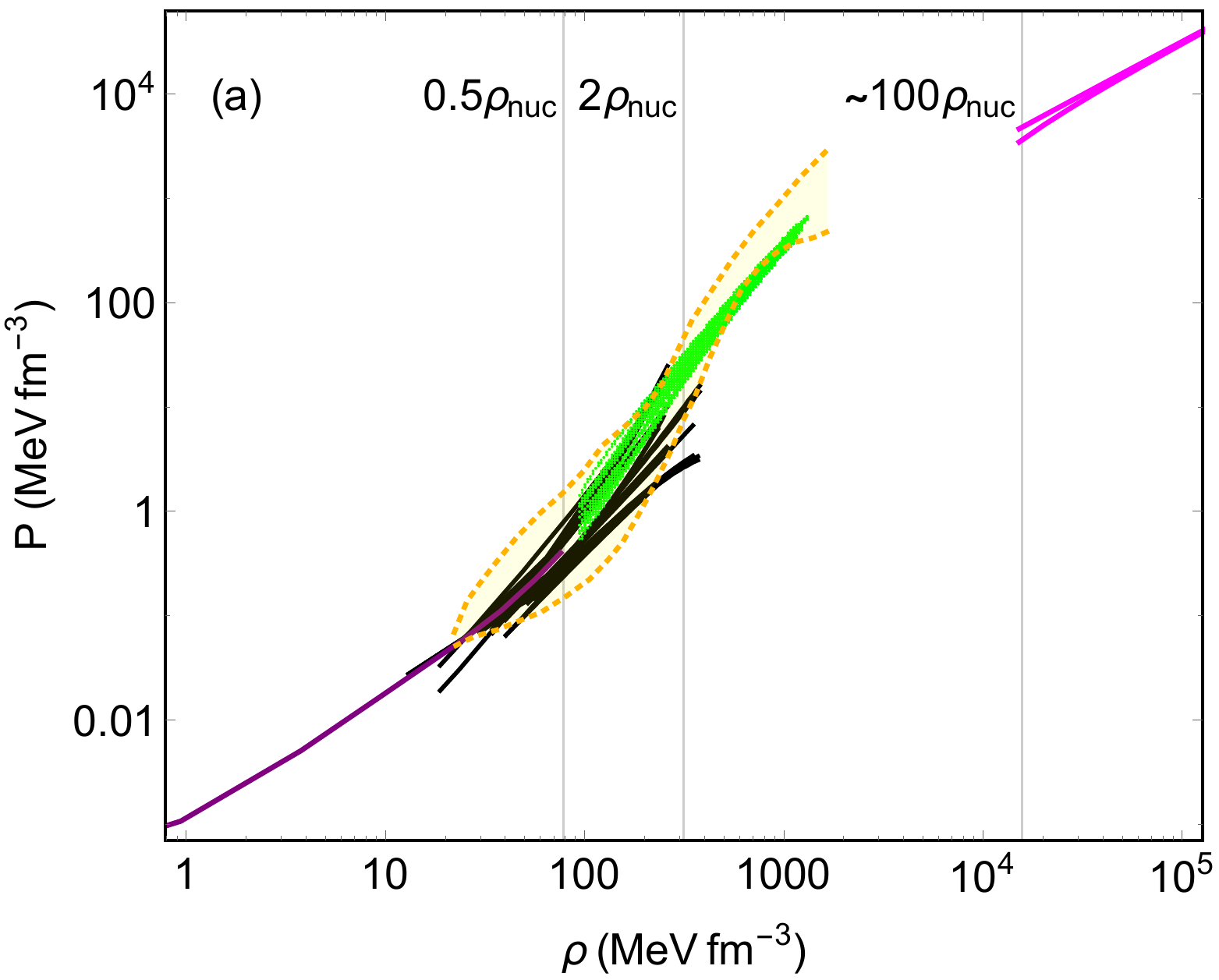}\quad
\includegraphics[width=7.8cm]{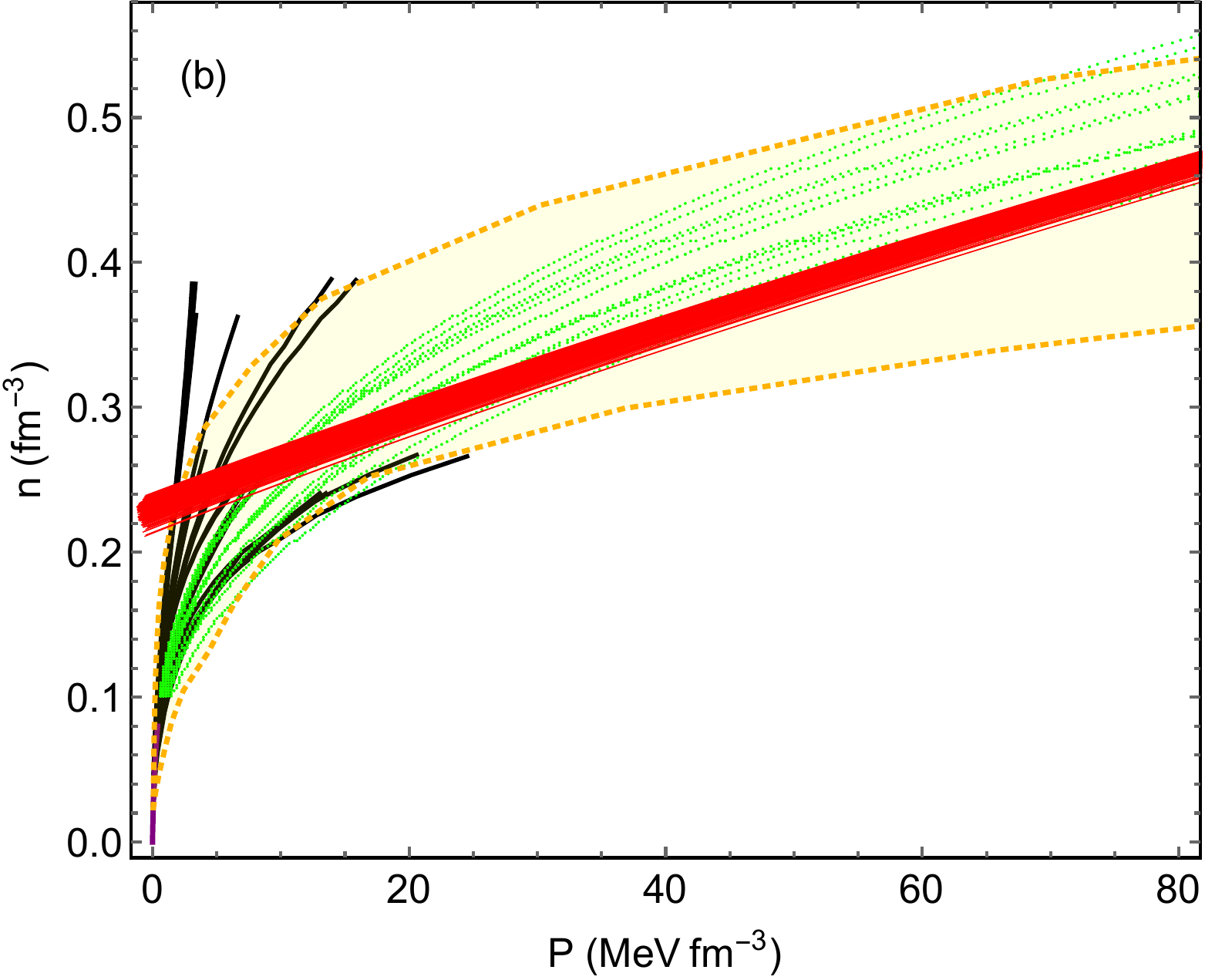}  
 \caption{(a) Hadronic matter EOSs on the $P-\rho$ plane; (b) hadronic matter and $ud$QM EOSs on the $n-P$ plane. The purple line shows a typical EOS for the crust. The black lines are predictions from the chiral perturbation theory (ChPT)~\cite{ucm_nEOS}. The green lines are some well known models for the $npe\mu$ fluid in the interior~\cite{Compose} that agree with the ChPT within uncertainties. The yellow band shows the joint constraints (the 90\% credible interval) on hadronic matter EOSs from the recent gravitational wave and pulsar observations~\cite{Landry:2020vaw}. The magenta lines in (a) denote the pQCD results. The red band in (b) is the prediction of $ud$QM with the $B_{\rm eff}$ range in Eq.~(\ref{Bbound1}). $\rho_{\rm nuc}=157\, \textrm{MeV}\,\textrm{fm}^{-3}$ is the nuclear saturation density.}
 \label{NSEOS}
\end{figure}

For a more comprehensive understanding of the conversion of neutron stars, we start from a brief review of the hadronic matter EOSs in compact stars. A typical neutron star has an atmosphere and an interior. Fig.~\ref{NSEOS}\,(a) summarizes our current understanding of the hadronic matter EOSs in the interior. Below $0.5\rho_{\rm nuc}$ is a curst consisting of ions and electrons (and free neutrons when the density is above  the neutron drip density). The curst EOS is testable in laboratory and is known to a good accuracy.
The outer core that ranges from $0.5 \rho_{\rm nuc}$ to $2\rho_{\rm nuc}$ is a mixture of protons, neutrons, electrons and sometimes muons in $\beta$-equilibrium. 
Its EOS has been systemically studied in the Chiral Perturbation Theory (ChPT) with baryons, and the theoretical uncertainties are well under control. 
A heavy neutron star may have an inner core with an intermediate density $\rho\gtrsim 2\rho_{\rm nuc}$. Although predictions are made by some models, including hyperons or not, the composition and EOS in this region remain largely unknown. 
Astrophysical observations for neutron stars provide important clue to the EOS in this region. Comparing with the joined constraints from recent observations, some models for example are disfavored at $\rho\gtrsim 4\rho_{\rm nuc}$.
Perturbative QCD (pQCD) applies at an ultrahigh density, i.e. $\rho\gtrsim 100\rho_{\rm nuc}$. Although this region is far from accessible in a neutron star, its prediction may serve as an asymptotical limit for any model of the intermediate density region. Phenomenologically, an EOS needs to satisfy the monotony and causality conditions, i.e. $0\leq dP/d\rho\leq 1$.

As highlighted in Sec.~\ref{sec:transition}, the spacing between the hadronic matter and $ud$QM $n(p)$ curves is crucial in determining the transition rate. Here we present the comparison of various $n(p)$ curves in Fig.~\ref{NSEOS}\,(b).  
Given the theoretical range of $B_{\rm eff}$ in Eq.~(\ref{eq:Beffrange}), the $ud$QM prediction is a quite narrow band as approximated by $n(p)\approx 0.003\textrm{MeV}^{-1} P+n(0)$.
Interestingly, most of the hadronic models considered before are quite soft, and their $n(p)$ curves can easily intersect with the $ud$QM band at some low pressure, i.e. below $30 \, \rm MeV\,fm^{-3}$, accessible from an astrophysical neutron star. Thus, newly formed hadronic stars described by these EOSs will experience an instantaneous transition, and observed compact stars with $M\gtrsim 1.4M_\odot$ are most likely to be $ud$QSs. 
On the other hand, the uncertainty range of the hadronic matter EOSs as from the low energy theory and astrophysical observations remain large, where the major part of the $ud$QM band is covered.  A slow transition is then possible for a special set of viable EOSs with the $n_H(P)$ curve sitting moderately below the $ud$QM band. 
In the following, we discuss these two possibilities and their observational implications in detail.

As a side remark, for the case that a crossing of $n(p)$ curves occurs, the chemical potential difference $\mu_H-\mu_Q$ starts to decrease above the crossing point with $n_H>n_Q$, referring to Eq.~(\ref{mup_int}).
At some higher pressure, $\mu_H$ may become smaller than $\mu_Q$, indicating that the hadronic matter becomes more stable again. If this pressure is accessible from $ud$QSs, there will be a transition back to hadronic matter in the deep interior of quark stars. This points to a new type of hybrid stars, in contrast to the conventional ones with a quark matter core. 
We leave the detailed study for future work.

\subsection{All compact stars being $ud$QSs}
\label{sec:scenario1}

Neutron stars described by a soft hadronic matter EOS is more likely to convert to $ud$QSs, the maximum mass of which remains compatible with the observed heaviest pulsars. The possibility that all compact stars are $ud$QSs then provides a natural solution to the hyperon puzzle.  
For this case, the main question is the consistency of $ud$QS predictions with most of the other neutron star observations that involve objects considerably lighter than $2M_\odot$.  Note that the joined constraints found in \cite{Landry:2020vaw} and other references rely on the nuclear theory input for hadronic matter at the low density, and they cannot be directly used for $ud$QM and $ud$QSs.  

\begin{figure}[h]
\centering
\includegraphics[width=10cm]{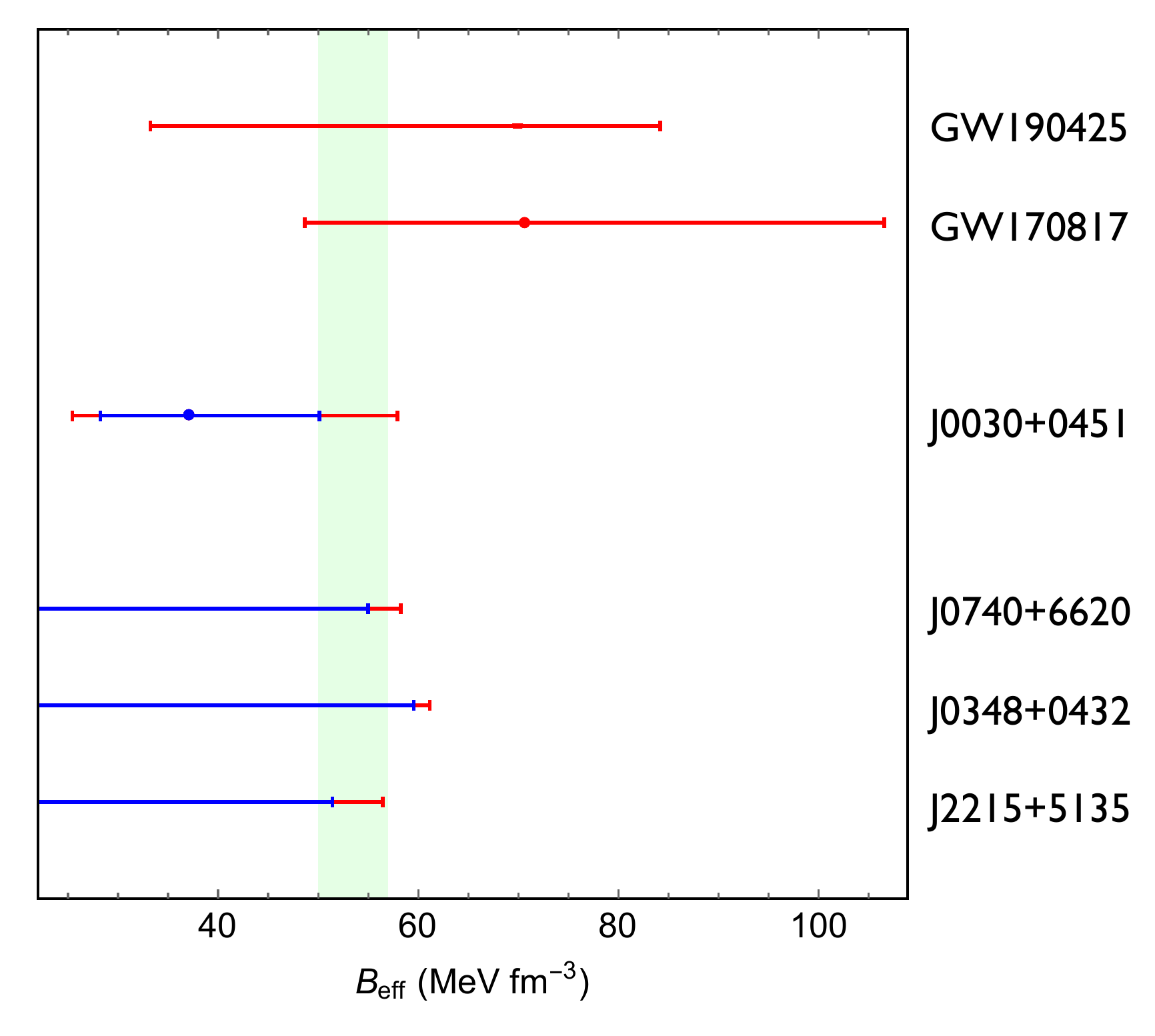}  
 \caption{Observational constraints on $B_{\rm eff}$ at 68\% C.L. (blue) and 90\% C.L. (red) in the scenario that all compact stars are $ud$QSs. The three constraints on the bottom show the upper bounds on $B_{\rm eff}$ from radio measurements of heaviest pulsar masses, with $M/M_{\odot}=2.14_{-0.09}^{+0.10} $, $2.01_{-0.04}^{+0.04} $, $2.27_{-0.15}^{+0.17}$ (68\% C.L.) for J0740+6620~\cite{Cromartie:2019kug}, J0348+0432\cite{Antoniadis:2013pzd}, J2215+5135\cite{Linares:2018ppq}, respectively. The constraint in the middle comes from the mass and radius measurement of J0030+0451 with NICER. Here we use the results from a recent analysis~\cite{Miller:2019cac} with $M=1.44^{+0.15}_{-0.14}M_\odot$ and $R=13.02^{+1.24}_{-1.06}\,$km (68\% C.L.). The two constraints on the top are from gravitational wave measurements of neutron star binaries with LIGO/Virgo, with the chirp mass $M_c/M_\odot=1.186\pm0.001$, $1.44\pm0.02$, the mass ratio $q=0.72\textrm{--}1$, 0.8--0.1, the average tidal deformability $\tilde\Lambda= 300_{-230}^{+420}$, $\lesssim 600$ (90\% C.L.) for GW170817~\cite{TheLIGOScientific:2017qsa} and GW190425~\cite{Abbott:2020uma} respectively. The vertical green band shows the theoretical prediction in Eq.~(\ref{Bbound1}).}
\label{fig:BeffconsS1}
\end{figure}

Figure \ref{fig:BeffconsS1} compares the theoretical range of $B_{\rm eff}$ in Eq.~(\ref{Bbound1}) with the recent gravitational wave and pulsar observations of neutron stars. Three types of constraints are considered here. 
For the observations of massive pulsars with $M\gtrsim 2M_\odot$, we include another pulsar J2215+5135 with a heavier mass but a much larger uncertainty. 
Overall, the theoretical range is consistent with these $2M_\odot$ bounds at 90\% C.L..  
NICER measures the X-ray emission from a rotating neutron star and is expected to reach better sensitivity for the mass and radius measurements. As the first target, J0030+0451 points to a star with the mass around $1.4M_\odot$ and the radius around $13\,$km. Given the rescaled mass and radius relation for $ud$QSs in Fig.~\ref{fig:$ud$QS}\,(a), the inferred range on the $M-R$ plane can be translated to a range for $B_{\rm eff}$. A relatively small $B_{\rm eff}$ is favored by this observation, with the theoretical prediction disfavored at 68\% C.L.. The tension nonetheless goes away at 90\% C.L., and there is even less concern if considering the theoretical uncertainties associated with the surface tension $\sigma_{s0}$. 

Gravitational wave observations provide a unique chance to measure the tidal properties for the binary system. The average tidal deformability $\tilde{\Lambda}$ can be extracted from a waveform at the inspiral stage, and it is a function of the mass ratio and the rescaled chirp mass. The constraints from GW170817 favor a relatively large $B_{\rm eff}$, with the theoretical prediction sitting right within the lower boundary of the 90\% range~\cite{Zhang:2019mqb}. The more recent event GW190425, on the other hand, has a larger chirp mass and imposes a much weaker lower bound on $B_{\rm eff}$. The upper bound comes solely from the observed mass for the heavier $ud$QS, where the rescaled $\bar{M}$ exceeds the maximum allowed value for a too large $B_{\rm eff}$. More details on tidal constraints can be found in Appendix~\ref{tidal}.  
As we can see, different observations push $B_{\rm eff}$ towards the opposite directions, while the theoretical prediction remains compatible with all the constraints at 90\% C.L..

On the other hand, there are a few neutron star observations that the compatibility with quark stars remains under debate. 
One long-established phenomenon is the pulsar glitch, a sudden increase of the pulsar spin frequency,  as being observed for the Vela and Crab pulsars. The most popular interpretation involves a superfluid component and a rigid structure~\cite{Anderson:1975zze,Alpar:1977}, with glitches produced by their angular momentum transfer. In the standard scenario, the rigid structure is provided by a solid crust, and the crustal moment of inertia is bounded from below by the observations of ``giant glitches". Although still under debate~\cite{Alpar:1987vk, Glendenning:1992kd, Haensel:2007yy}, the normal nuclear curst of a quark star below the neutron dip pressure~\cite{Alcock:1986hz} is probably too small to account for the demanded crustal momentum inertia. An alternative for quark stars is a crust consisting of small chunks of quark matter instead of ordinary nuclei. For strange stars, this new crust might be large enough with the energy density contributed mainly by the strangelets~\cite{Jaikumar:2005ne}. The two ingredients may also be related to the peculiar properties of quark matter. One example is the inhomogeneous crystalline color superconducting phase, which is rigid as well as superfluid and may provide an explanation without a crust~\cite{Anglani:2013gfu}.  
For the stable $ud$QM case, the nuclear crust for $ud$QS would be larger than that for strange stars due to a larger positive charge for $ud$QM and a stronger Coulomb support of the crust. Implications of other mechanisms deserve further studies. 

A more recent example for such kind of observations is quasi-periodic oscillations for the highly magnetized compact stars. In the simplest model, they are associated with the seismic oscillations of the stellar crust, and the frequencies are determined mainly by the crust thickness. Quark stars are disfavored due to their much thinner crust and the much higher frequency, even considering a crust consisting of quark matter~\cite{Watts:2006hk}. However, to infer the crust thickness, modes identification between the observation and theory is needed, and this may depend crucially on other unknown features of the stars~\cite{Miller:2018kmk}. 
All in all, the current observations in tension with the quark star explanation seem to involve complicated structure of quark matter, and further studies are required for a more conclusive analysis.

Mergers of quark stars may produce small chunks of $ud$QM, which we name as $ud$lets, in line with the strangelets in the SQM hypothesis. Normally, a $ud$let would not be absorbed by an ordinary nucleus due to the Coulomb repulsion of the positive charges. But those generated from mergers may acquire large kinetic energy to overcome the Coulomb barrier, and their encounter with smaller hadronic stars, e.g. white dwarfs, planets, may lead to fast conversion into small quark stars. The final results depend on the flux and spectrum of $ud$lets. 
A recent numerical simulation~\cite{Bauswein:2008gx} shows that the strangelet flux from strange stars merger is negatively correlated with $B_{\rm eff}$ through the mass-radius relation. Since $ud$QM has a smaller $B_{\rm eff}$ than SQM with the same $\varepsilon_{\rm min}$, the conversion rate induced by the $ud$lets might be higher than that for strangelets.

\subsection{Co-existence of hadronic stars and $ud$QSs}
\label{sec:scenario2}

In the two-families scenario, high-mass stars with $M\sim 2M_\odot$ are all $ud$QSs, while low-mass ones remain hadronic with a slow enough transition rate. As mentioned before, this points to a special class of hadronic matter EOSs, which is a little fine-tuned. But in view of observations, it shows that the transition behavior in this scenario is extremely sensitive to the variations of hadronic matter EOSs, and could be used to provide information that is otherwise inaccessible. Another advantage of this scenario is the possibility to avoid the long-time debate regarding the compatibility with observations such as pulsar glitches, given that these observations are consistent with lower mass stars within uncertainties.

\begin{figure}[h]
\centering
\includegraphics[width=10cm]{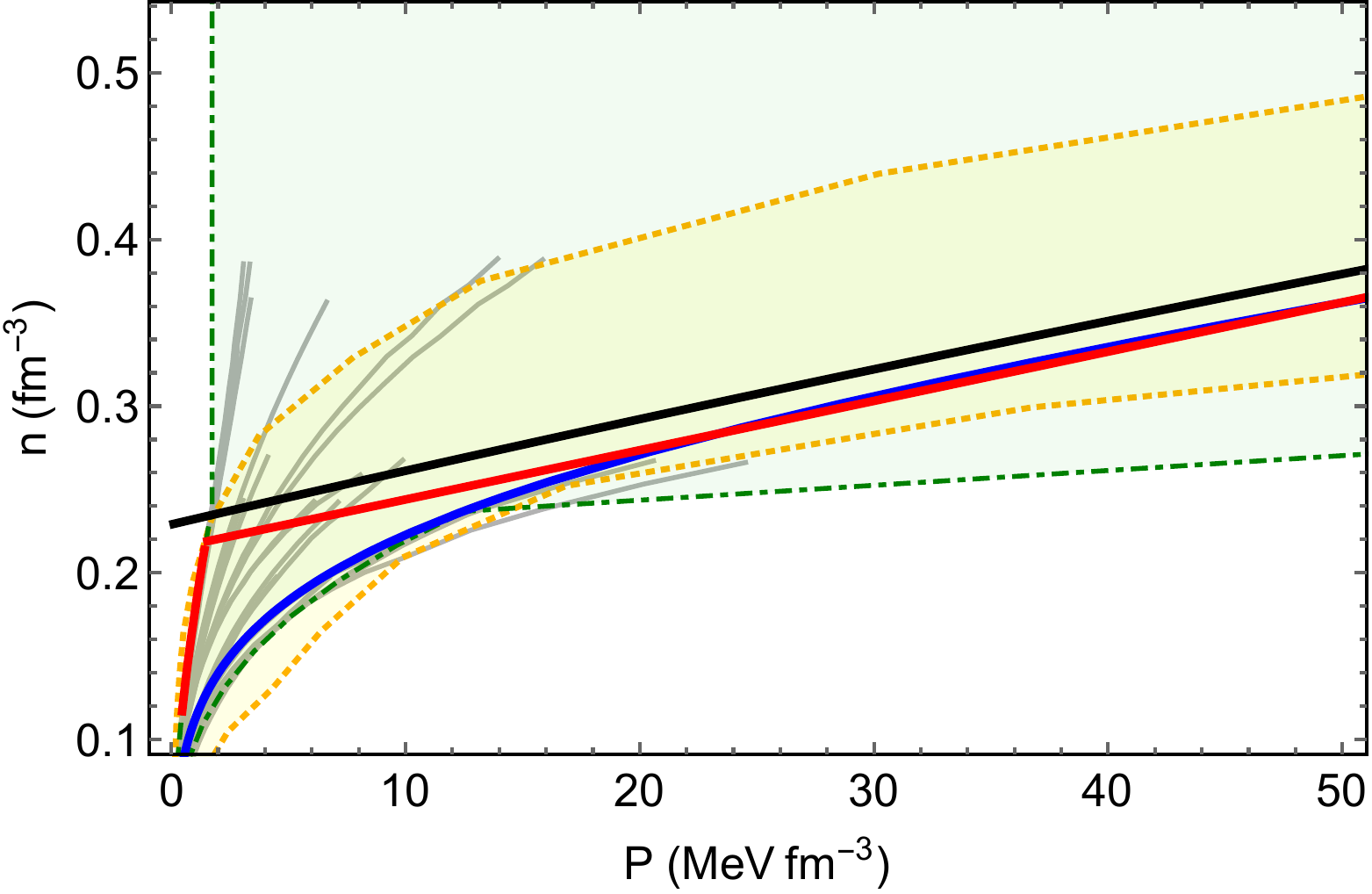}  
 \caption{The hadronic matter and $ud$QM EOSs on the $n-P$ plane. The gray lines are predictions from the ChPT~\cite{ucm_nEOS}. The yellow band (dotted line) shows joint constraints (the 90\% credible interval) from astrophysical observations~\cite{Landry:2020vaw}. The green band (dotdashed lines) shows the range of EOS that the low density ChPT results and high density pQCD results can be interpolated under the causality and monotony condition, i.e. $0\leq dP/d\rho\leq 1$.  The black line is the prediction of $ud$QM with $B_{\rm eff}=52\,\rm MeV\,fm^{-3}$. The blue and red lines are the two benchmark models of the hadronic matter EOSs, HM1 and HM2, as detailed in the text.}
 \label{NSEOS2}
\end{figure}

Figure~\ref{NSEOS2} displays two benchmark models for hadronic matter EOSs on the $n-P$ plane. To avoid an instantaneous transition, the hadronic matter $n_H(P)$ curve is bounded from above by the $n_Q(P)$ curve. On the other hand, it cannot be too stiff and is bounded from below by the observational constraints as well as the requirement of matching to the high-density pQCD prediction. 
For EOS in this range, the center pressure $P_c$ for $1.4M_{\odot}$ neutron stars varies only in a narrow range, i.e. $P_c\approx 35\textrm{--}45\,\rm MeV\,fm^{-3}$. Thus, only the behavior at $P\lesssim 50\,\rm MeV\,fm^{-3}$ is relevant to transition of $1.4M_{\odot}$ neutron stars, as we focus on in Fig.~\ref{NSEOS2}.  
The blue line denotes an example (HM1) that smoothly interpolates the high density and low density regimes as described by the pQCD and ChPT. Matching to the crust at $0.5\rho_{\rm nuc}$, its EOS at higher density is given by the analytical expression\footnote{We thank Bob Holdom for providing a preliminary version of this expression.}
\be
n(P) = (a + b P) (1 - \text{exp}(-c P^d))-f,
\label{nP_Bob}
\ee 
with the parameters $(a,b,c,d,f)=(0.4,0.0019,0.4,0.3,0.04)$. Below $20\,\rm MeV\,fm^{-3}$, this model is close to the stiffest EOS within the ChPT uncertainty band. 
The red line, on the other hand, is roughly the softest EOS (HM2) allowed by the ChPT that is below the $ud$QM curve. The abrupt change of the slope at $P\approx 1.5\,\rm MeV\,fm^{-3}$ indicates a drastic variation of the speed of sound, which may come from a phase transitions inside neutron stars.

\begin{figure}[h]
\centering
\includegraphics[width=7.8cm]{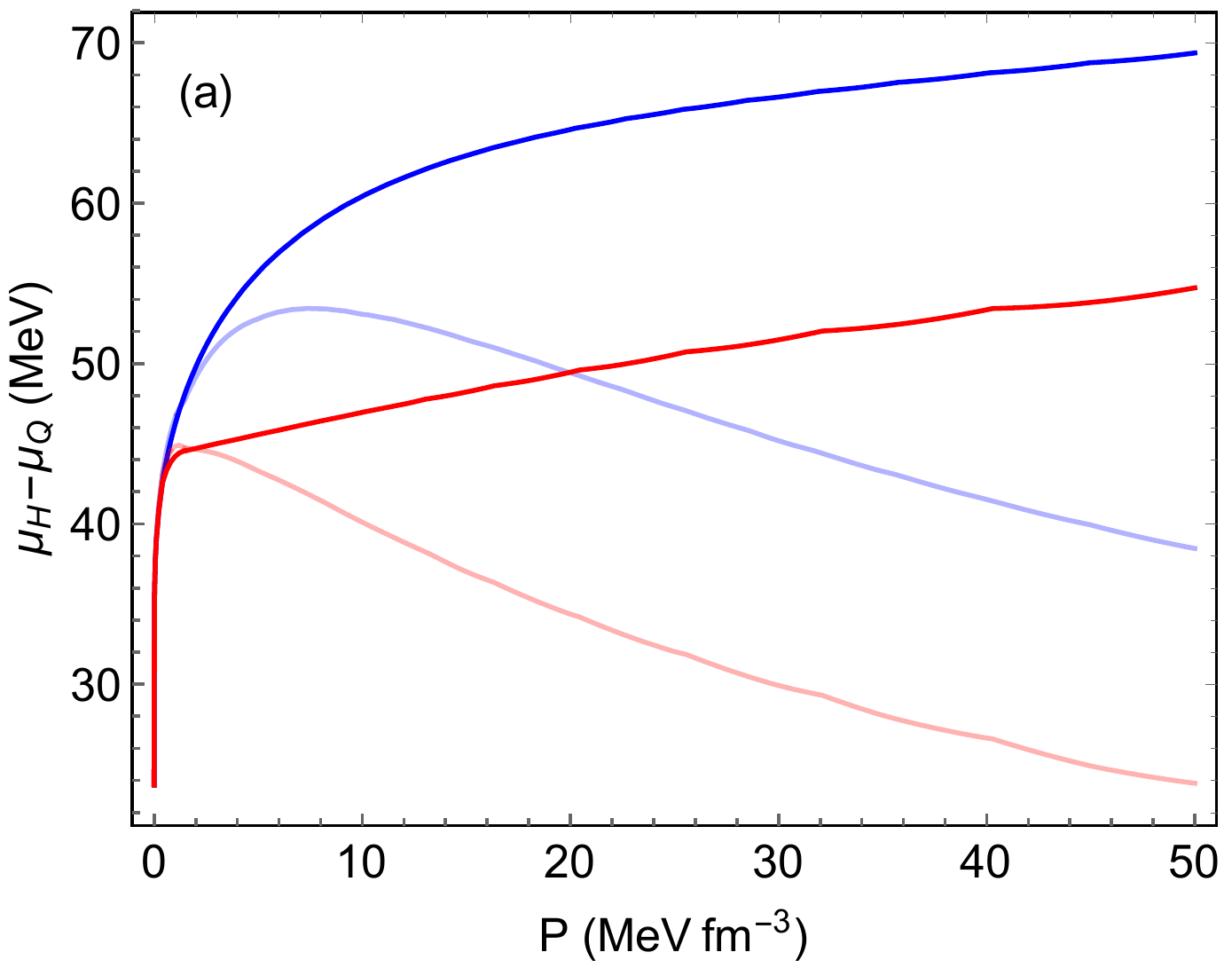}\quad
\includegraphics[width=8cm]{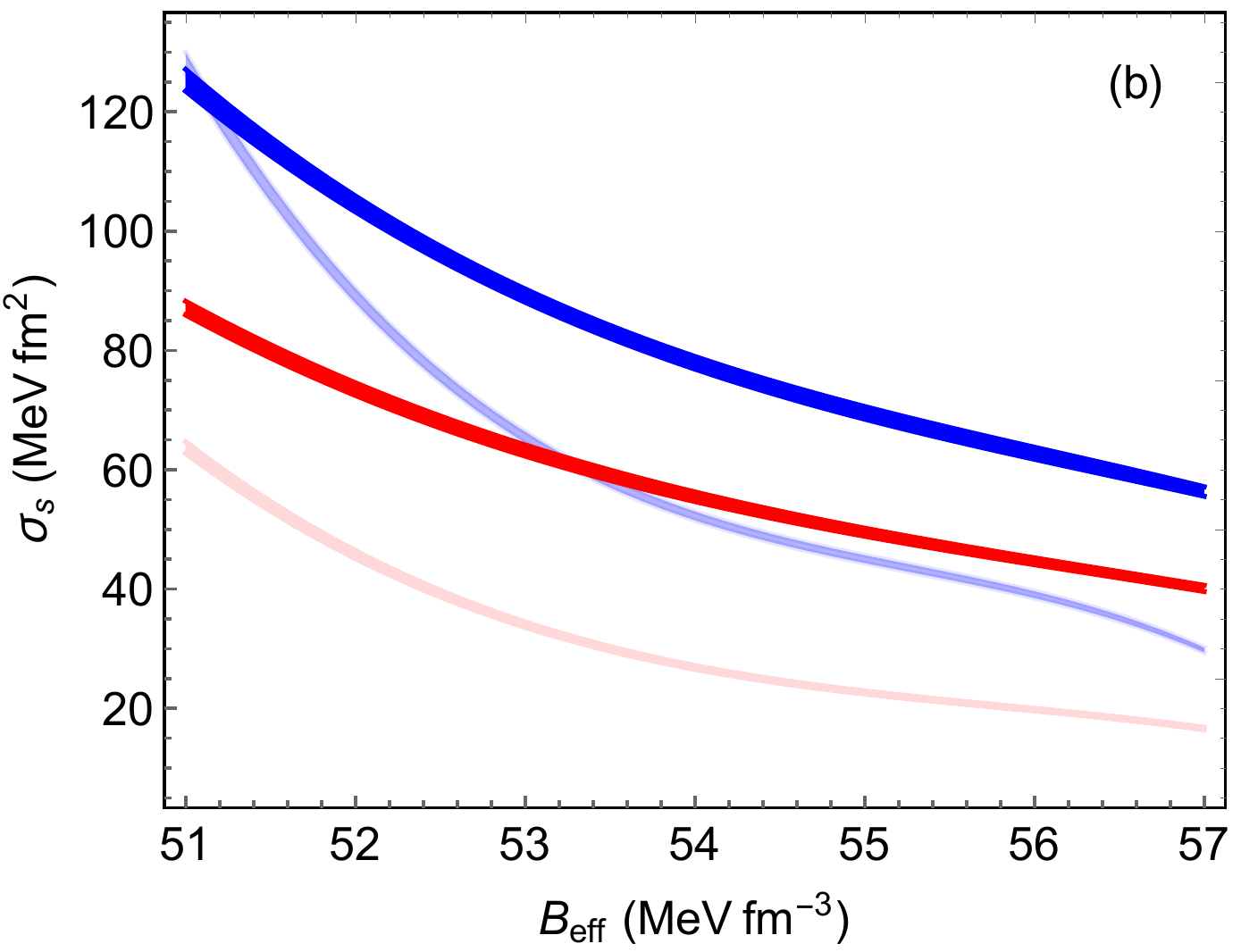}  
 \caption{(a) Chemical potential difference as a function of the pressure for $B_{\rm eff}=52\,\rm MeV\,fm^{-3}$. (b) Contours of the transition time $\tau_s=t_0$ for 1.4$M_{\odot}$ neutron stars on the $(\sigma_s, B_{\rm eff})$ plane for $N_s\approx 10^{45}$--$10^{55}$, where we restrict to the range of $B_{\rm eff}$ in Eq.~(\ref{eq:Beffrange}). On both panels, the blue and red lines correspond to HM1 and HM2 in Fig.~\ref{NSEOS2}. The dark and light lines are for $ud$QM with a negligible lepton fraction and with a large lepton fraction from the Sk15 model~\cite{Compose}. }
 \label{fig:transition}
\end{figure}    

Figure \ref{fig:transition} shows the main properties relevant to the transition of 1.4$M_{\odot}$ neutron stars. Different benchmark models of hadronic matter and $ud$QM are chosen to demonstrate the dependence on the effective bag constant $B_{\rm eff}$, the surface tension $\sigma_s$, the hadronic matter EOS and the flavor composition. As shown in Fig.~\ref{fig:Cscan}, the transition time is determined by the competition among $C_M$, $C_P$ and $\sigma_s$. 
A large $\sigma_s$ and $B_{\rm eff}$ both raise the height of the potential barrier and increase the transition time.  
For the two hadronic matter models HM1 and HM2, their $n(P)$ approach the same values at high density, and the difference of the transition time in Fig.~\ref{fig:transition}\,(b) mainly comes from the distinct $C_P$, as being proportional to the chemical potential difference in Fig.~\ref{fig:transition}\,(a). Since HM2 is softer at lower density, it has smaller $\mu_H$ and then a smaller chemical potential difference.

A nontrivial $P$-dependence of the flavor composition can also be helpful by reducing the chemical potential difference at high pressure. For illustration, we focus on hadronic matter models with a negligible strangeness, and the flavor composition varies mainly with the lepton fractions.\footnote{There are larger uncertainties for EOSs involving hyperons. On the quark matter side, a nonzero strange fraction will increase $\mu_Q$ and a slow transition can be more easily achieved.}  For the case with negligible contribution from leptons, given the thermodynamic relation Eq.~(\ref{mup_int}), the chemical potential difference $\mu_H-\mu_Q$ increases with pressure as expected from the condition $n_H(P)<n_Q(P)$ , and the transition at the center of stars is the fastest. In the presence of nonzero lepton fractions, the chemical potential of $ud$QM becomes larger, and the difference $\mu_H-\mu_Q$ can be significantly reduced at high pressure due to the nontrivial $P$-dependence of the lepton fractions, which reach up to 20\% for this model.

It is then clear that a soft hadronic matter EOS at lower density and a nontrivial lepton fractions can help to slow down the transition. For instance, as from Fig.~\ref{fig:transition}\,(b), assuming HM2 and a large lepton faction, it is possible to have 1.4$M_\odot$ hadronic stars not yet converted at present for the $B_{\rm eff}$ range in Eq.~(\ref{eq:Beffrange}) and for the surface tension as small as our prediction for the quark-vacuum interface, i.e. $\sigma_{s0}\approx 20\,\textrm{MeV}\,\textrm{fm}^{-2}$.
The narrow bands of the transition time contours in Fig.~\ref{fig:transition}\,(b) results from the exponential sensitivity of $\tau_sN_s$ on the shape of the potential barrier. A small shift of parameters within the band leads to a change of $\tau_s N_s$ by ten orders of magnitude, corresponding to a variation of the transition time $\tau_s$ from the age of the universe to one year for $N_s\approx 10^{55}$.  Thus,  in the parameter space above or below the bands, the transition time of a 1.4$M_\odot$ hadronic star is either too long or too short in the time scale of interest to human beings.

For heavy neutron stars with $M\approx 2M_{\odot}$, the stiffness of hadronic matter EOS is essential for the discussion of transition. For a stiff EOS that the maximum mass is already compatible with the observed heaviest pulsars, e.g. $M_{\rm max}\approx 2.1M_{\odot}$ for HM1, the conventional one-family picture has no direct conflict with observations, and a fast transition of heavy hadronic stars is not mandatory. It is for a soft EOS with a smaller $M_{\rm max}$ that a fast transition of heavy hadronic stars to $ud$QSs is motivated.
As a soft EOS in the two-families scenario, HM2 has $M_{\rm max}\approx 2M_{\odot}$ and $P_c\approx 100\text{--}200\,\textrm{MeV}\,\textrm{fm}^{-3}$ for $M\approx 1.9\text{--}2M_{\odot}$. The transition time of heavy hadronic stars then depend on the EOS in a large range of pressure above the central pressure $P_c\approx 50\,\textrm{MeV}\,\textrm{fm}^{-3}$ for $1.4M_\odot$ stars. A fast transition can be realized if $n_H$ gets close to $n_Q$ or $\mu_H-\mu_Q$ becomes large enough at any $P$ within this range, and these conditions are easy to achieve given the theoretical uncertainties in Fig.~\ref{NSEOS2}\,(a). For HM2 with negligible leptons, a $1.9M_{\odot}$ hadronic star can convert to a $ud$QS almost instantaneously in the range above the dark red band covered by Fig.~\ref{fig:transition}\,(b). The viable parameter space might not be this large for other cases. But it is clear that a fast transition of heavy stars are very likely.  

Regarding the implication of recent gravitational wave and pulsar observations, different comparisons need to be made in the two-families scenario. Observations for the heaviest pulsars still constrain $ud$QSs, and the upper bounds on $B_{\rm eff}$ remain the same as in Fig.~\ref{fig:BeffconsS1}. 
The NICER observation of a pulsar with $M\approx 1.4 M_\odot$, on the other hand, is to be confronted with the prediction of hadronic stars. For the two benchmarks HM1 and HM2, the corresponding radii, which are $13.3,\, 12.5$\,km, are quite compatible with the NICER results.   
The implication of gravitational wave observations are also different if the binaries involve at least one hadronic star. For such cases, as shown in Fig.~\ref{fig:QSHS}, the hadronic matter EOS plays an important role, with quite different results for the two benchmarks. HM1 is ruled out by GW170817 at 90\% C.L. for either a $ud$QS-HS system or a HS-HS system simply due to a too large tidal effect for the hadronic star. The situation for HM2 is better, where we find no constraints for the HS-HS case and $B_{\rm eff}\gtrsim 50 \rm MeV\,fm^{-3}$ for the $ud$QS-HS case.

In the two-families scenario, low mass hadronic stars not yet converted by the quantum nucleation may experience a fast transition by encounter with a $ud$let, which can be produced by binary mergers involving heavy $ud$QSs. 
Recent numerical simulations for strange stars show that the merger product of quark stars tends to promptly collapse to a black hole with much less ejecta~\cite{Bauswein:2008gx}. 
A binary with heavy $ud$QSs either have a too small companion or a too large total mass. For the former case the merger is too mild to produce ejecta, while for the latter case matter is mostly swallowed by the promptly formed black hole. Thus, the $ud$let flux as coming only from binary mergers involving heavy $ud$QSs would be much smaller than the flux in the ``all compact stars being $ud$QSs" case, and the chance of low-mass stars converted by $ud$lets is expected to be small.

\section{Summary}
\label{sec:conc}

We investigated astrophysical implications of the stable $ud$QM scenario in this paper, taking into account both the transition rate estimation for hadronic stars and the observational constraints. 

With the effective bag constant $B_{\rm eff}$ range (\ref{eq:Beffrange}) derived from the $ud$QM stability condition~\cite{HRZ2017}, we found the predicted maximum mass of $ud$QSs compatible with observations of the heaviest pulsars with $M\gtrsim 2M_\odot$ at 90\% C.L.. Therefore, after a fast conversion to $ud$QSs, heavy hadronic stars lighter than $2M_\odot$ can keep growing till saturating the maximum mass of $ud$QSs, and this provides a natural solution to the hyperon puzzle. 

The main issue we addressed here is the nature of low-mass compact stars with $M\sim 1.4M_\odot$. As shown in Fig.~\ref{fig:Cscan}, the transition time mainly depends on the chemical potential and the density difference for the hadronic matter and quark matter phases, as well as the surface tension $\sigma_s$ of their interface. We found it convenient to track the EOS dependence through comparison of the $n(P)$ curve of the two phases. As a result, the transition rate only becomes significant when moving into the interior. A prominent feature is that when the two $n(P)$ curves cross, the $ud$QM droplets turn ultra-relativistic, and the transition is instantaneously fast regardless of the values for other quantities.

Most of the hadronic models do predict a $n(P)$ curve intersecting with that of $ud$QM at a pressure accessible from a $1.4M_\odot$ compact star. This then points to the unconventional possibility that the observed neutron stars are mostly $ud$QSs. This possibility is often overlooked due to a long-time debate on its compatibility with some well-established observations. Yet, complicated structures of quark stars are likely to be involved. For a more direct probe of the basic properties of $ud$QM, we consider constraints from the recent gravitational wave and pulsar observations. As shown in Fig.~\ref{fig:BeffconsS1}, different observations push $B_{\rm eff}$ to the opposite directions with a small region left open. We found the theoretical prediction of $ud$QM still viable at 90\% C.L., which may resurrect interest in this possibility.

A slow transition of low-mass hadronic stars is also possible if the hadronic matter $n(P)$ curve happens to be moderately below the $ud$QM one, which is still allowed given the uncertainties. 
For this case, the transition time is extremely sensitive to variations of the relevant  quantities, as summarized in Fig.~\ref{fig:transition}. 
We found that a softer hadronic matter EOS at low pressure and a nontrivial lepton fraction can help to slow down the transition, and a reasonable lower bound on $B_{\rm eff}$ and $\sigma_s$ can be obtained to have $1.4M_\odot$ compact stars being hadronic at the present universe.
Heavy hadronic stars with $M\approx 2M_\odot$, on the other hand, can quickly convert to quark stars.
Thus, the transition behavior in the two-families scenario provides useful information for both  $ud$QM and hadronic matter. 
A softer hadronic EOS is favored by the recent observations as well, in particular GW170817 from LIGO/Virgo. 

There are more to explore in the future. On the theoretical side, further model development for stable $ud$QM may help to limit the allowed ranges for $B_{\rm eff}$ and $\sigma_{s}$, which will lend to a more definite conclusion for the two-families scenario.  
On the observational side, a hadronic star conversion is a dramatic event, where a large amount of energy is expected to be released. This may trigger a neutrino burst accompanied by emission of gravitational waves~\cite{Bombaci:2016xuj}. The implication for $ud$QSs deserves further studies.

\begin{acknowledgments}

We thank Bob Holdom for early collaboration and valuable discussions.
J.R. is supported in part by the Institute of High Energy Physics, Chinese Academy of Sciences, under Contract No. Y9291120K2. C.Z. is supported in part by the Natural Sciences and Engineering Research Council of Canada.

\end{acknowledgments}	

\appendix
\section{Tidal deformability of $ud$QSs}
\label{tidal}
In this section, we discuss in detail the tidal deformability constraints from the compact star merger events on $ud$QSs, and we use the geometric unit with $G=c=\hbar=k_B=1$ here.

The GW170817 event detected by LIGO/Virgo~\cite{TheLIGOScientific:2017qsa} is the first confirmed merger event of compact stars, with  the chirp mass $M_c=1.186^{+0.001}_{-0.001} \, M_{\odot}$, and a $90\%$ highest posterior density interval of $ \tilde{\Lambda}=300^{+420}_{-230}$ with $q=0.73-1.00$ for the low spin prior from the collaboration~\cite{TheLIGOScientific:2017qsa, Abbott:2018wiz}.  More recently, a new event GW190425 was identified~\cite{Abbott:2020uma} with $M_c=1.44\, M_\odot$, $q=0.8-1.0$ and $ \tilde{\Lambda} \leq 600$ for the low spin prior at 90\% credible interval.
Ref.~\cite{Zhang:2019mqb} showed that the GW170817 event may be a binary system with at least one $ud$QS.  In the following, we update the constraints in the context of the neutron star conversion, and extend the discussion to GW190425.

\begin{figure}[h]
  \centering
    \includegraphics[width=7.9cm]{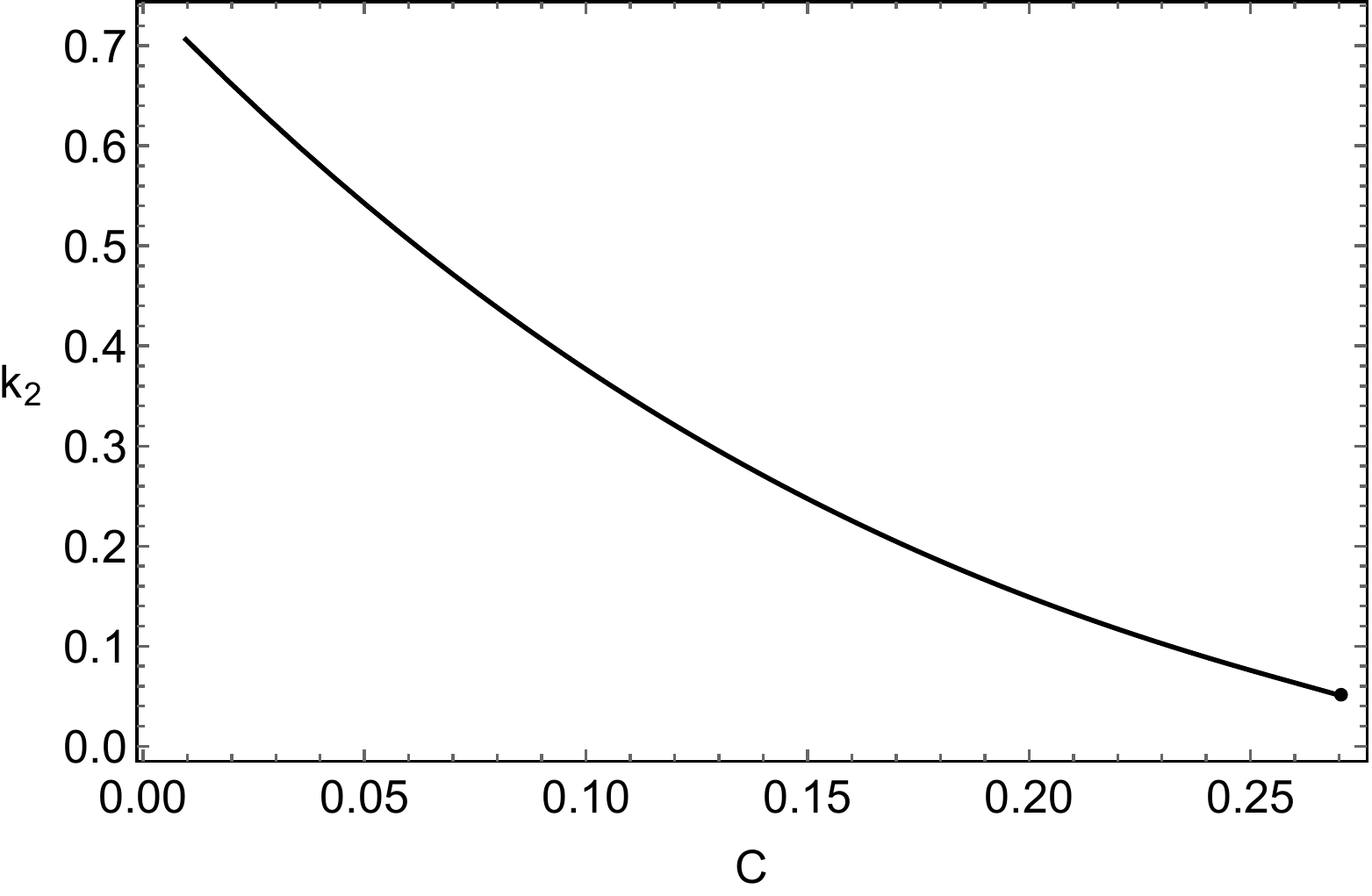}  
        \includegraphics[width=7.9cm]{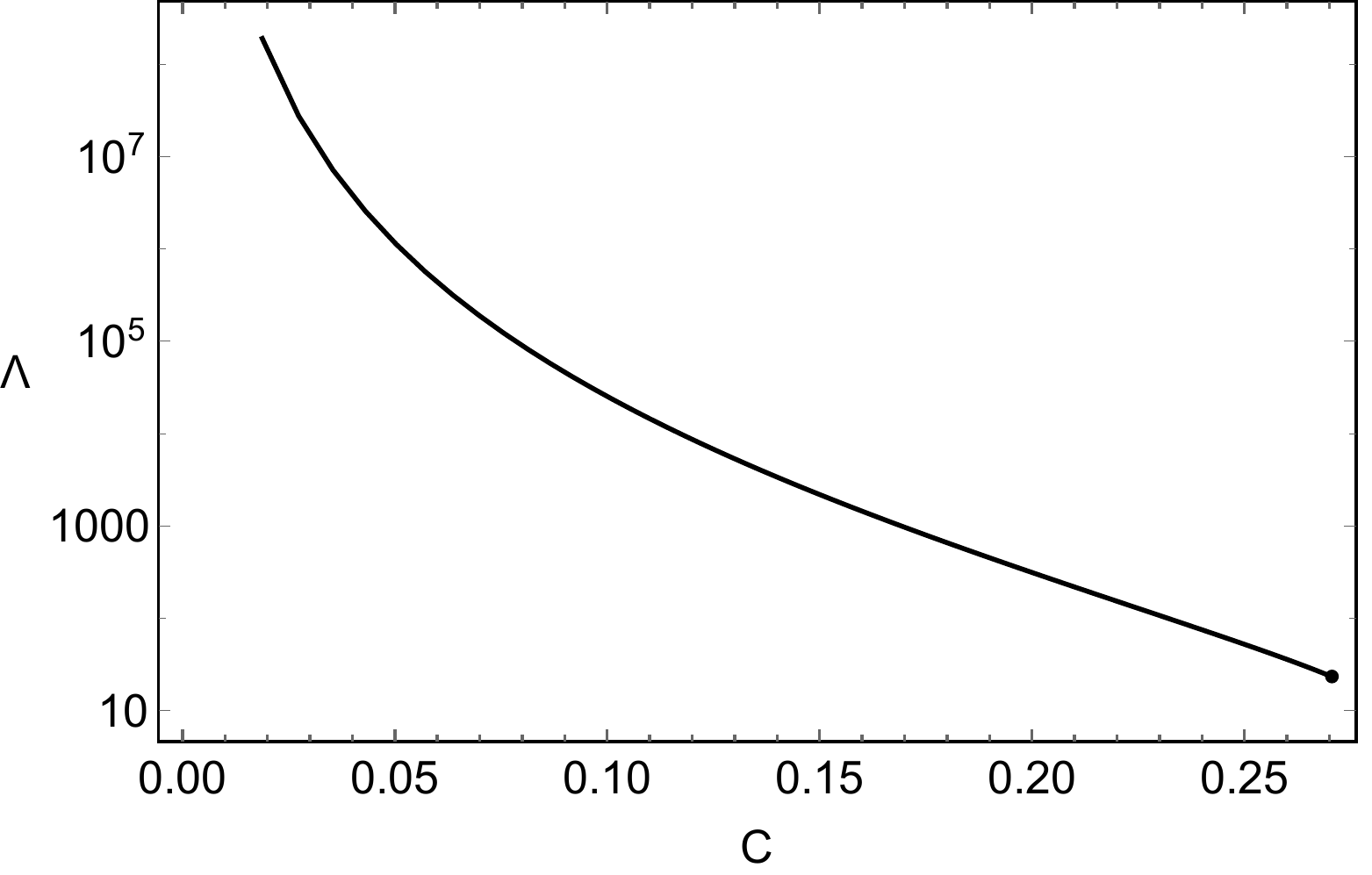}  
   \caption{ (a) Love number $k_2$ and (b) tidal deformability $\Lambda$ as functions of the compactness $C=M/R=\bar{M}/\bar{R}$ for $ud$QSs.}
  \label{fig:k2C}
\end{figure}

The tidal deformability $\Lambda$, which characterizes the response of compact stars to an external disturbance, can be expressed as $\Lambda=2k_2/(3C^5)$. The Love number $k_2$ is defined as~\cite{AELove,Hinderer:2007mb,Hinderer:2009ca,Postnikov:2010yn}
\bea
\begin{aligned}
k_2 &= \frac{8 C^5}{5} (1-2C)^2 [ 2 + 2 C (y_R-1) -y_R ] \times \{ 2 C [ 6- 3y_R + 3C (5y_R-8)]\\
&+4C^3[ 13-11 y_R+ C (3y_R-2)+2C^2(1+y_R)]  \\
&+ 3 (1-2C)^2 [ 2 - y_R + 2C (y_R-1)] \log (1-2C )\}^{-1}~,
\label{eqn:k2}
\end{aligned}
\eea   
where the compactness $C=M/R$, and $y_R$ is $y(r)$ evaluated at the surface, which can be obtained by solving the following equation \cite{Postnikov:2010yn}:
\bea
\begin{aligned}
&ry^\prime(r)+y(r)^2+ r^2Q(r) +y(r)e^{\lambda(r)}\left[1+4\pi r^2(P(r)-\rho(r))\right]=0\,,
\label{eqn:y}
\end{aligned}
\eea
with boundary condition $y(0)=2$. Here
\bea
\begin{aligned}
Q(r)&=4\pi e^{\lambda(r)} \left(5\rho(r)+9P(r)+\frac{\rho(r)+P(r)}{c_s^2(r)}\right)-6\frac{e^{\lambda(r)}}{r^2}-\left(\nu^\prime(r)\right)^2,
\label{eq:Q}
\end{aligned}
\eea
and 
\begin{equation}
e^{\lambda(r)}=\left[1-{2m(r)\over r}\right]^{-1}\,,\,
\nu^\prime(r)=2e^{\lambda(r)}{m(r)+4\pi P(r)r^3\over r^2}.
\label{eq:met}
\end{equation}
For quark stars with a finite surface density, a matching condition should be imposed at the boundary $y_R^{\rm ext}=y_R^{\rm int}-  4\pi R^3\rho_0/M$~\cite{Damour:2009vw}.  Note that we can also utilize Eq.~(\ref{rescale}) to transform Eq.~(\ref{eqn:y}) into a fully dimensionless form, with $\bar{\rho}$ and $\bar{P}$ obtained from the rescaled TOV equation for quark stars as introduced in Sec.~\ref{sec:QMQS}. The solution then is in the form of $y(\bar{r})$ with $\bar{r}=r\sqrt{4 B_{\rm eff}}$, and the variable $y(\bar{R})$ evaluated at the surface can be converted further into the $y(C)$ form with the $\bar{M}-\bar{R}$ relation in Fig.\ref{fig:$ud$QS}\,(a). Therefore,  for $ud$QSs, the Love number $k_2$ and the tidal deformability $\Lambda$ are only functions of the compactness $C$, as shown in Fig.~\ref{fig:k2C}\,(a), (b) respectively,
with the dependence on $\rho_0$ or $B_{\rm eff}$ fully absorbed. 
This feature crucially relies on the linear form of quark matter EOSs.

For a binary system, the average tidal deformability is defined as
\bea
 \tilde \Lambda &=& \frac{16}{13} 
\frac{ (1+12q)}{(1+q)^5} {\Lambda} (M_1)+ \frac{16}{13}  \frac{q^4 (12+q)}{(1+q)^5}{\Lambda}({M}_2)\,,
\label{LamLam}
\eea
where $q=M_2/M_1\leq 1$. For an equal mass binary with $q=1$, $\tilde \Lambda$ is simply $\Lambda(M_i)$. In the other limit that $q\to 0$, $\tilde \Lambda$ is dominated by the massive component contribution. 

\begin{figure}[h]
  \centering
       \includegraphics[width=10cm]{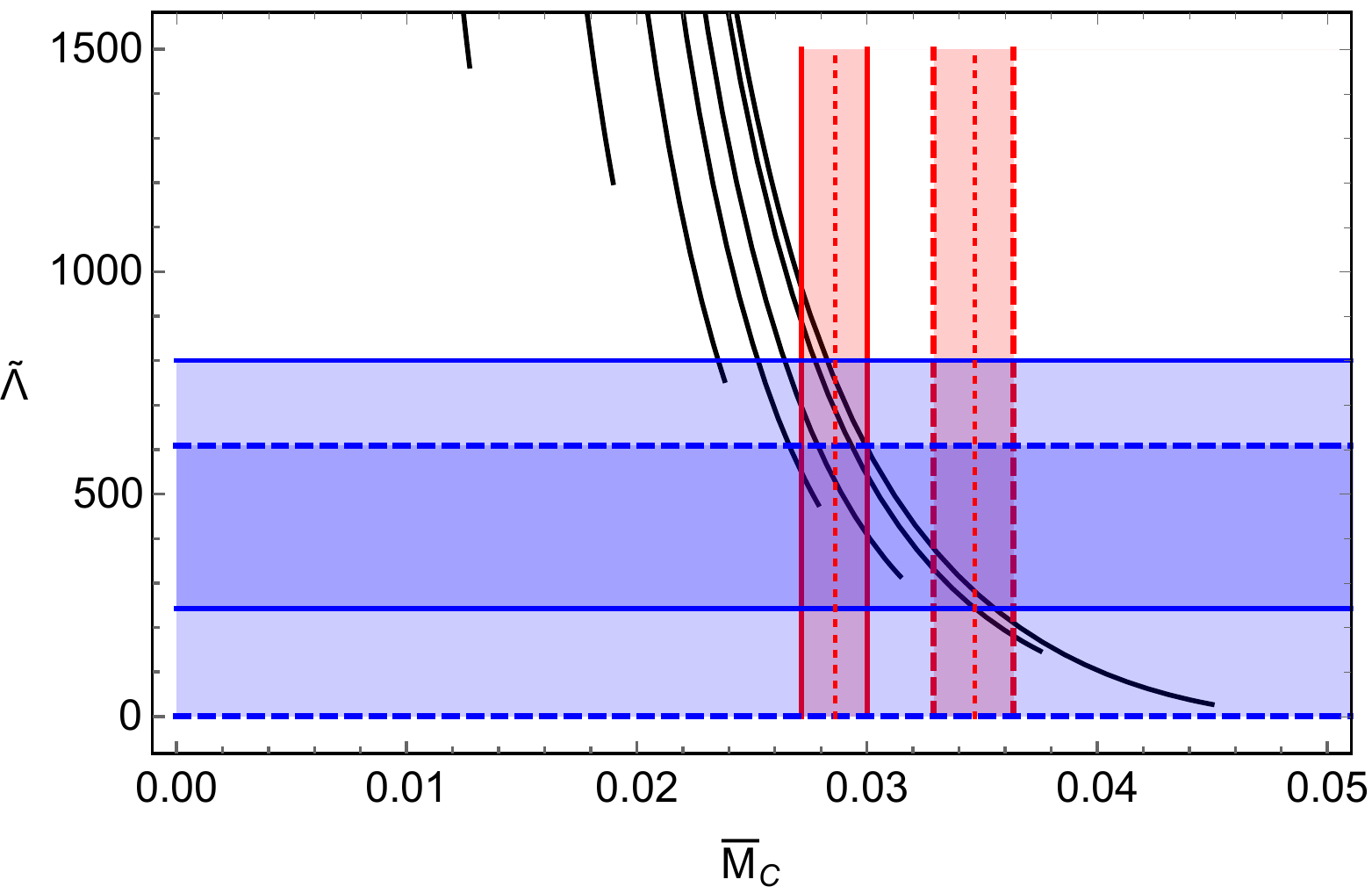}  
 \caption{The average tidal deformability $\tilde{\Lambda}$ vs the rescaled chirp mass $\bar{M}_c$ for the $ud$QS-$ud$QS merger case (black lines) for $q=(0.1, 0.2, 0.3,0.4, 0.5, 0.7, 1)$ from left to right, respectively. Red and blue bands with solid (dashed) edge lines denote the range of $\bar{M}_c$ with $B_{\rm eff}  \in [45 , \,55] \rm \, MeV\, fm^{-3}$ and the constraints on  $\tilde{\Lambda}$ for GW170817 (GW190425) respectively. The red dotted lines at the center of each red band denote $\bar{M}_c$ with $B_{\rm eff} = 50 \rm \, MeV\, fm^{-3}$.}
 \label{fig:QSQS}
  \end{figure}

For the $ud$QS-$ud$QS merger case,  with $\Lambda(M_i)=\Lambda(\bar{M}_i)$, the average tidal deformability is only a function of the mass ratio $q$ and the rescaled chirp mass $\bar{M}_c$.
The dependence is shown in Fig.~\ref{fig:QSQS}, where the lower end of each black curve for a given $q$ is determined by requiring each component of the binary system to not exceed its maximum allowed value. The value of  $\bar{M}_c$ at the lower end is negatively correlated with the value of  $q$.
  
We can see that GW170817 imposes a stronger constraint than GW190425. For both cases, the upper bound of the averge tidal deformability $\tilde{\Lambda}$ can be converted to a lower bound of $\bar{M_c}$ and thus a lower bound for $B_{\rm eff}$, and we find $B_{\rm eff}  \gtrsim 49.5, \,36.2 \rm \, MeV\, fm^{-3}$ for GW170817 and GW190425 respectively. Similarly, the lower bound of $\tilde{\Lambda}$ for GW170817 is translated to a quite mild upper bound $B_{\rm eff} \lesssim 106.6  \rm \, MeV\, fm^{-3}$. For GW190425, there is an upper bound by requiring the rescaled $\bar{M}$ to not exceed the maximum allowed value $\bar{M}_{\rm max}\approx0.052$, and this gives $B_{\rm eff}\lesssim84.2 \rm \, MeV\, fm^{-3}$ for $M_c=1.44\, M_{\odot}$. These bounds of $B_{\rm eff}$ map to the top two lines in Fig.~\ref{fig:BeffconsS1}.

\begin{figure}[h]
  \centering
       \includegraphics[width=8.15cm]{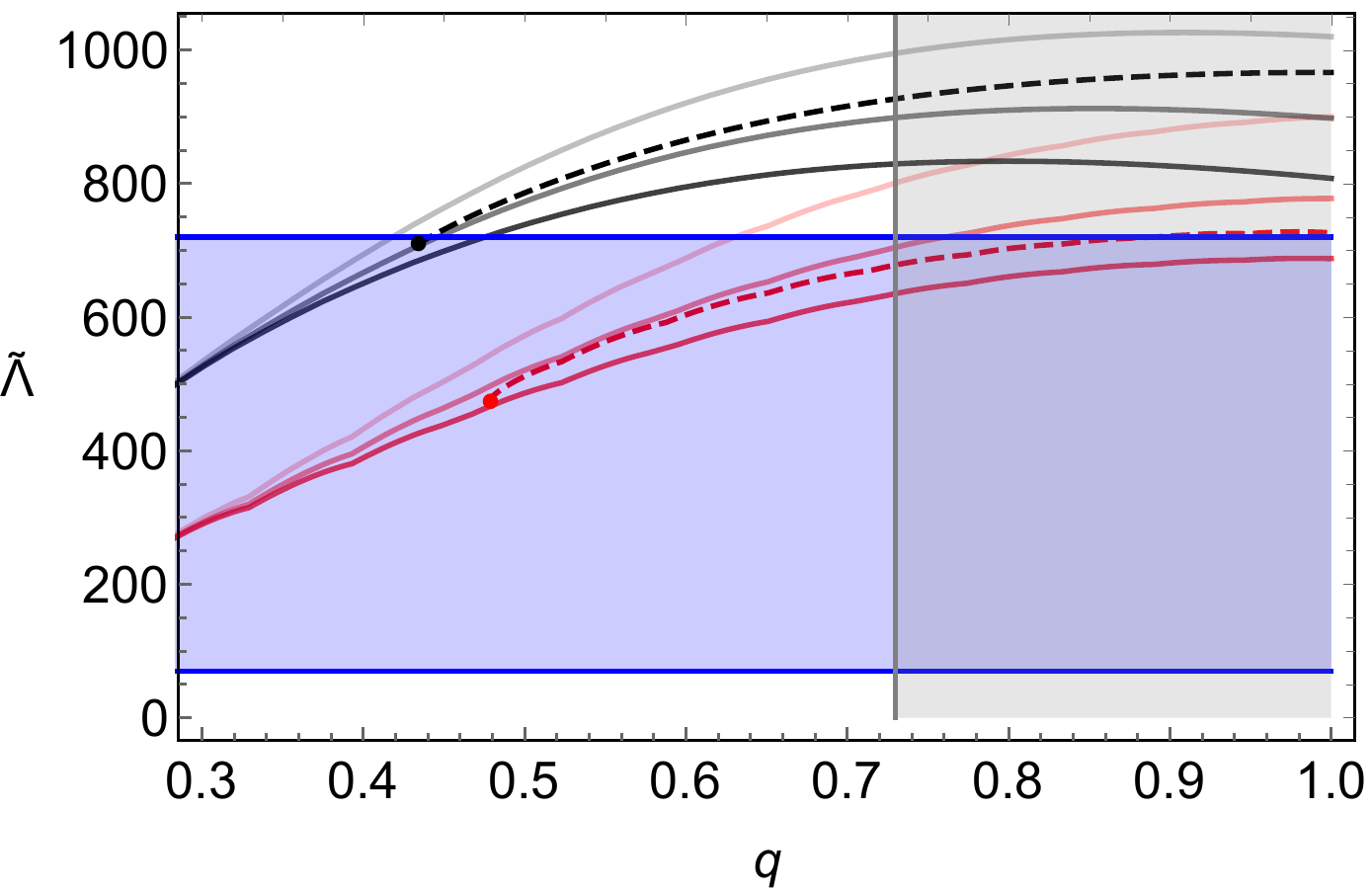} 
       \includegraphics[width=8.13cm]{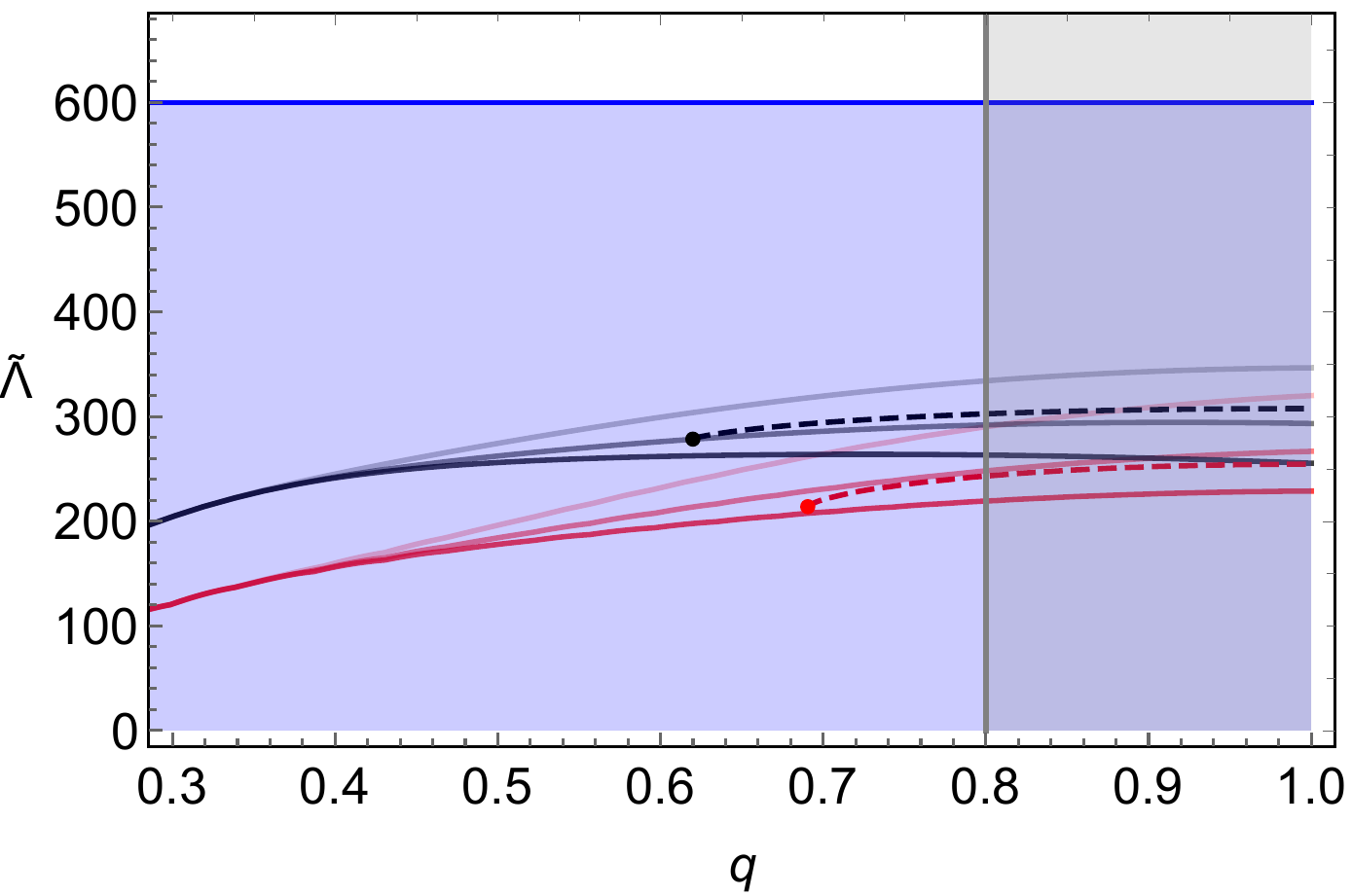} 
 \caption{The average tidal deformability $\tilde{\Lambda}$ vs $q$ for the two-families scenario for (a) GW170817 and (b) GW190425, where the binaries include at least one hadronic star. The blue and grey bands show the constraints on $\tilde{\Lambda}$ and $q$ from LIGO/Virgo. The black and red lines are for the two benchmark models of hadronic matter EOSs, HM1 and HM2, respectively. For each color, the dashed line denotes the HS-HS case, and the solid dots denote the lower bound on $q$ that ensures the component mass to not exceed the maximum allowed value. The solid lines are for the $ud$QS-HS case with $M_2$ the hadronic star mass and $B_{\rm eff}=(45, 50, 55){\rm\, MeV\, fm^{-3}}$ (a darker color for a larger value) for $ud$QSs.}
 \label{fig:QSHS}
  \end{figure}

For the two-families scenario, a binary with at least one low-mass star could be either a $ud$QS-HS system or a HS-HS system. Here we use the two benchmarks of hadron matter EOSs introduced earlier in Fig.~\ref{NSEOS2}, i.e. HM1 and HM2, which are proposed to realize a slow transition of the low-mass hadronic stars. This is in contrast to the previous study in Ref.~\cite{Zhang:2019mqb}, where Bsk19, SLy, Bsk21 models are chosen for a more general representation of the hadronic matter EOSs. 

The corresponding results of $\tilde \Lambda$ for GW170817 and GW190425 are shown in Fig.~\ref{fig:QSHS}. We can see that for GW190425 either a $ud$QS-HS merger or a HS-HS merger is well compatible with the current constraint. For GW170817, the observations favor a relatively soft hadronic EOS and $B_{\rm eff}\gtrsim 50 \rm MeV\,fm^{-3}$ for $ud$QM, which match the expectation of Ref.~\cite{Zhang:2019mqb}.

\end{document}